\documentclass[12pt]{article}
\usepackage[hmargin=2cm,vmargin=1.5cm]{geometry}
\usepackage{graphicx}
\usepackage{amsmath}
\usepackage{enumerate}

\begin{document}
\title{Quantum temporal probabilities in tunneling systems}
\author {Charis Anastopoulos\footnote{anastop@physics.upatras.gr} and   Ntina Savvidou\footnote{ksavvidou@physics.upatras.gr}\\
 {\small Department of Physics, University of Patras, 26500 Greece} }

\maketitle

\begin{abstract}
 We study the temporal aspects of quantum tunneling as manifested in time-of-arrival experiments in which the detected particle tunnels through a potential barrier. In particular, we present a general method for constructing temporal probabilities in tunneling systems that   (i) defines `classical' time observables  for quantum systems and  (ii) applies to relativistic particles interacting through quantum fields. We show that the relevant probabilities are defined in terms of specific correlations functions of the quantum field associated with tunneling particles. We construct a probability distribution with respect to the time of particle detection that   contains all information about the temporal aspects of the tunneling process. In specific cases, this probability distribution leads to the definition of a delay time that, for parity-symmetric potentials, reduces to the phase time of Bohm and Wigner. We apply our results to piecewise constant potentials, by deriving the appropriate junction conditions on the  points of discontinuity. For the double square potential, in particular, we demonstrate the existence of (at least) two physically relevant time parameters, the delay time and a decay rate that describes the escape of particles trapped in the inter-barrier region. Finally, we propose a resolution to the paradox of apparent superluminal velocities for tunneling particles.  We demonstrate that the idea of faster-than-light speeds in tunneling follows from an inadmissible use of classical reasoning in the description of quantum systems.
\end{abstract}

\section{Introduction}
The issue of the time that a   quantum particle takes to tunnel through a potential barrier has been studied since the early days of quantum mechanics \cite{Con, McC}. The search for an answer to this question has led to several different candidates for the tunneling time---for reviews, see Ref. \cite{rev1}---rather than to a single expression derived unambiguously from first principles. Many existing definitions imply that tunneling times saturate in the opaque-barrier limit, thus suggesting superluminal speeds for particles traversing sufficiently long barriers, a phenomenon known as the Hartmann effect \cite{Har62}.

In this paper, we identify  tunneling time through its association with quantum {\em temporal} observables,  i.e, time variables whose value can be determined in specific experiments. A quantum temporal observable is a random variable, so its determination requires the construction of a probability density rather than the specification of a single number. Hence, any temporal parameters that characterize the tunneling process must be identified in terms of a probability distribution associated with physical measurements on the tunneling particles. In particular, we propose a definition of tunneling time  in terms of the time instant that a detector, located far from the barrier, records tunneling particles, i.e.,   a definition  in terms of time-of-arrival measurements.
In our opinion, this  is the most natural operational definition of tunneling time, because a remote detector does not interfere with the tunneling process---unlike other proposed definitions of tunneling time in terms of external actions on the barrier region \cite{Larmor, Bula}.

A  time-of-arrival experiment typically involves a particle source and a particle detector separated by distance $L$ at rest with respect to another. Source and detector are equipped with a pair of synchronized clocks. The time of arrival is   the difference $t$ between the clock readings of detection and emission respectively. This `classical' definition of the time of arrival can be extended to quantum particles using the Quantum Temporal Probabilities (QTP) method \cite{AnSav12}. The main difference is that  for quantum particles, the time difference $t$ is a random variable that may take different values in different runs of the experiment. Thus, a quantum description involves the construction of a probability density $P(L,t)$ for the time of arrival. When a potential barrier is placed along the line between emitter and detector, all information about the temporal aspects of tunneling (the tunneling time and any other physically relevant parameter) is contained in the probability density $P(L, t)$ \cite{AnSav08}--- for other approaches to tunneling time in terms of time-of-arrival measurements, see Refs. \cite{AHN, HSMN, KRW}.

The construction of time-of-arrival probabilities associated with tunneling systems is a non-trivial task. In fact, the definition of time-of-arrival probabilities has run into ambiguities analogous to the ones in the definition of tunneling time---see, the reviews  \cite{ML00, timeqm} and references therein. The difficulty originates from the special role of time in quantum mechanics, where the time $t$ appearing in Schr\"odinger's equation is an external parameter and not as an ordinary observable, like position or momentum.  As a result, the squared modulus of the time-evolved wave-function, $|\psi(x, t)|^2$, is not a density with respect to $t$, and, hence, it cannot serve as a definition for the required probabilities.

\subsection{Our approach to the tunneling-time problem}

 In this work, we employ a general method for the construction of quantum temporal probabilities associated with any experimental configuration, that was developed in Ref. \cite{AnSav12}.
 The QTP method incorporates the detector degrees of freedom into the quantum description, so that the temporal probabilities are always defined with respect to specific experimental set-ups. The key  idea is to distinguish between the roles of time as a parameter to Schr\"odinger's equation and as a label of the causal ordering of events \cite{Sav99, Sav10}. This important distinction leads to the definition of quantum temporal observables. In particular, we identify the time of a detection event as a coarse-grained quasi-classical variable \cite{GeHa93, hartlelo} associated with macroscopic records of observation. The time variables in QTP correspond to macroscopic observable magnitudes, such as the coincidence of a detector `click' with the reading of a clock external to the system.

A key property of the QTP method is that it applies to {\em any quantum system}, including relativistic quantum fields. Besides time-of-arrival probabilities, the method has been applied to the modeling of particle detectors in high-energy processes \cite{AnSav12}, and to the study of temporal correlations in accelerated detectors \cite{AnSav11}. Earlier versions of QTP \cite{AnSav06} have been employed in studies of non-relativistic tunneling times \cite{AnSav08}, non-exponential decays \cite{An08} and time-extended measurements \cite{AnSav07}.

   Using the QTP method, we derive a probability density $P(L, t)$ describing time-of-arrival measurements on relativistic particles that tunnel through a potential barrier. The interactions of the particles with the barrier and the detector are described in terms of  Quantum Field Theory (QFT).

The probability density $P(L,t)$  derived here incorporates all temporal aspects of the tunneling process. For a specific class of  potentials and initial states, the only timescale relevant to tunneling is a temporal variable $t_d$.  This variable describes the delay in the transit time of particles crossing through the barrier in comparison to the transit time   in absence of the barrier. For parity-symmetric potential, $t_d$ coincides with the Bohm-Wigner phase time \cite{BW}, obtained from the asymptotic analysis of wave-packets. However, our identification of the delay time does not employ non quantum-invariant notions, such as the peak of the wave packet or its center of mass, but follows from the structure of the probability density $P(L,t)$, in specific regimes \cite{asymp}.

 In general, a single time parameter, such as $t_d$, does not capture all temporal aspects of the tunneling process---for the different time-scales in quantum tunneling see, Ref. \cite{BM94}. We demonstrate this point by studying tunneling through a double square barrier. We find that the probability distribution $P(L,t)$ involves two distinct timescales, a delay time $t_d$, as described earlier, but also a decay time $\Gamma^{-1}$ that determines the escape rate of particles trapped between the two barriers. Previous work on this system, led by the {\em a priori} assumption of a single-time scale governing tunneling, had led to a very different physical description of the tunneling process, according to which the tunneling time is largely independent of the inter-barrier distance \cite{genhar, genhar2}.

\subsection{Proposed resolution of the superluminality paradox}
Based on the results described above, we propose a new resolution to the superluminality paradox in tunneling. This paradox originates from the fact that many existing definitions of tunneling time imply that the tunneling time saturates in the opaque-barrier limit. This suggests superluminal speeds for particles traversing sufficiently long barriers (the Hartmann effect \cite{Har}).

Several experimenters have reported superluminal tunneling velocities in electromagnetic analogues of quantum tunneling \cite{EN, SKC, SSSK, BD, MSHM, LMLB}. The analogy is based on the mathematical correspondence between the classical Helmholtz equation for the electromagnetic field in inhomogeneous dielectric media and the time-independent Schr\"odinger equation in presence of a potential. The experiments above measure the group delay $t_d$ associated with electromagnetic pulses crossing a barrier of length $d$. In classical field theory, the group delay is standardly interpreted as the transit time of a signal through a medium or a device. Hence, the group velocity $d/t_d$ (which is found greater than the speed of light) is interpreted as the transit velocity for waves crossing the barrier. However, no direct measurement of the transit time is involved in these experiments; the transit time is inferred from the group delay.

Here, we point out that the relation between group delay and transit time is a feature of classical physics and not of quantum physics. In quantum theory, there is no such relation, because group delay and transit time are incompatible observables that are determined in distinct types of experiment. The former is an observable in experiments that measure the wave aspects of a quantum system, while the latter refers to the propagation of localized excitations of quantum fields, i.e., particles. The interpretation of group delay as a transit time violates a fundamental rule of quantum mechanics, Bohr's complementarity principle, according to which "evidence obtained under different experimental conditions cannot be comprehended within a single picture" \cite{Bohr}.

Group delay in electromagnetic analogues of tunneling has a natural interpretation as the lifetime of energy stored in the barrier \cite{Win}. To address the issue of superluminality, it is necessary to consider experimental set-ups in which the particles'  transit time (or time of arrival) is directly observed, as we do in this paper. The probability density $P(L,t)$, derived here, contains all information about signal propagation in tunneling and it is defined in terms of
QFT (rather than relativistic quantum mechanics \cite{rel}). The latter feature is essential for a resolution of the superluminality paradox, because superluminal speeds imply a violation of the principle of local causality. This principle is implemented in quantum theory only if the interactions are expressed in terms of local quantum fields \cite{Wein, Haag}. Indeed, we show that the propagation of signals in tunneling systems is fully causal, because the QTP method leads to a definition of $P(L,t)$  in terms of the correlation functions of a local quantum field theory.

\subsection{Main results}

The main results of the paper are the following.
\begin{enumerate}{}
\item We present a general methodology for defining the time-of-arrival probabilities $P(L,t)$ in tunneling set-ups that is valid for any potential barrier and for any method of particle detection.     We derive the probability distribution $P(L,t)$ for relativistic particles interacting through quantum fields. We show that $P(L,t)$ is obtained from a suitable two-point function of the quantum field smeared with a kernel that depends on the physics of the detection process.

\item For the standard case of a potential barrier model by a background classical field, we derive a simple expression for the probability distribution $P(L,t)$ as a positive linear functional of the initial state.  The probability distribution $P(L,t)$
    can be explicitly constructed from the knowledge of the transmission and reflection amplitudes of the potential.

\item We find that the total transmission probability coincides with the square modulus of the transmission amplitude associated with the potential only if the potential is parity symmetric. In general, they differ. This is because the QFT propagator, through which the probability distribution $P(L,t)$ is defined, involves contributions from both left-moving and right-moving eigenstates of the single-particle Hamiltonian.

\item We construct the probability distribution $P(L,t)$ explicitly for piecewise constant potentials, by deriving the appropriate junction conditions on the  points of discontinuity of the potential. For the double square potential, in particular, we demonstrate explicitly that a single time parameter does not suffice to capture all information about the temporal aspects of tunneling.

\item We propose a resolution of the superluminality paradox in tunneling. We show that $P(L,t)$ is constructed from the correlation functions of a local QFT, and thus it cannot lead to superluminal signals.
\end{enumerate}

The structure of the paper is the following. In Sec. 2, we formulate the quantum tunneling of particles through a barrier in a quantum field theory language. In Sec. 3, we derive a general expression for the probability density $P(L, t)$ of detection of particles through a potential barrier. In Sec. 4, we construct explicitly the probability density $P(L, t)$ for piecewise-constant potentials. In Sec. 5, we present our proposal resolution to the superluminality paradox. In Sec. 6, we summarize our results and discuss further applications.

\section{QFT formulation of tunneling}
In this section, we formulate particle tunneling through a potential barrier in a quantum field theory language. In particular, we consider  a complex scalar field $\phi(x)$, corresponding to particles of mass $m$ and charge $e$,  interacting with a background  static electric field. The electromagnetic potential is $A_{\mu}(x) = (A_0(x), 0)$, where $A_0(x)$ differs from zero only in a spatial region $D$. For simplicity, we restrict our considerations to one spatial dimension, and we choose the origin of the coordinate system so that $D = [-d/2, d/2]$, where $d$ is the length of the barrier.

The classical Lagrangian density for the scalar field $\phi(x)$ is
\begin{eqnarray}
{\cal L}(\phi, \phi^*) = |\dot{\phi} - i V \phi|^2 - |\partial_x \phi|^2 - m^2 |\phi|^2,
\end{eqnarray}
where  $V(x) = e A_0(x)$.

The associated Hamiltonian density ${\cal H}$ is
\begin{eqnarray}
{\cal H} = |\pi|^2 + |\partial_x \phi|^2 + m^2 |\phi|^2 + i V (\pi \phi - \pi^* \phi^*), \label{hamden}
\end{eqnarray}
where $\pi$ and $\pi^*$ are conjugate field variables to $\phi$ and $\phi^*$ respectively.

In order to quantize the field, we first introduce two copies $H_1$ and $\bar{H}_1$ of the Hilbert space ${\cal L}^2(R, dx)$ of square integrable functions on the real line. $H_1$ and $\bar{H}_1$ are associated with a single particle and  with a single antiparticle, respectively.  The Hilbert space of the field is defined as ${\cal F}(H_1) \otimes {\cal F}(\bar{H}_1)$. ${\cal F}(H)$ stands for  the bosonic Fock space associated with the Hilbert space $H$,
\begin{eqnarray}
{\cal F}(H) = C \oplus H \oplus (H \otimes H)_S \oplus (H \otimes H \otimes H)_S \oplus \ldots ,
\end{eqnarray}
where $S$ denotes symmetrization.

The creation and annihilation operators $\hat{a}(x)$ and $\hat{a}^{\dagger}(x)$ for  particles and $\hat{b}(x)$ and $\hat{b}^{\dagger}(x)$ for  anti-particles are standardly defined on the Fock space ${\cal F}(H_1) \otimes {\cal F}(\bar{H}_1)$. They satisfy the   commutation relations

\begin{eqnarray}
[\hat{a}(x), \hat{a}^{\dagger}(x')] = \delta (x - x') \hspace{2cm}
[\hat{b}(x), \hat{b}^{\dagger}(x')] = \delta (x - x'),
\end{eqnarray}
with all other commutators vanishing.

The creation and annihilation operators are also expressed in their smeared form
\begin{eqnarray}
\hat{a}(f) = \int dx \hat{a}(x) f(x), \hspace{1cm} \hat{a}^{\dagger}(f) = \int dx \hat{a}^{\dagger}(x) f(x)\\
\hat{b}(f) = \int dx \hat{b}(x) f(x), \hspace{1cm} \hat{b}^{\dagger}(f) = \int dx \hat{b}^{\dagger}(x) f(x),
\end{eqnarray}
where $f$ is  a square-integrable function on $R$.

In order to represent the Hamiltonian Eq. (\ref{hamden}) as an operator on the Fock space ${\cal F}(H_1) \otimes {\cal F}(\bar{H}_1)$ we introduce the
 field operators $\hat{\phi}(x), \hat{\phi}^{\dagger}(x)$ and their conjugate momenta $\hat{\pi}(x), \hat{\pi}^{\dagger}(x)$. The field operators are defined in their smeared form as

\begin{eqnarray}
\hat{\phi}(f) := \frac{1}{\sqrt{2}} \left( \hat{a}(h_0^{-1/2}f) + \hat{b}^{\dagger}(h_0^{-1/2} f)\right) \hspace{1cm}
\hat{\pi}(f) := \frac{1}{\sqrt{2}i} \left( \hat{a}(h_0^{1/2}f) - \hat{a}^{\dagger}(h_0^{1/2} f)\right) \\
\hat{\phi}^{\dagger}(f) := \frac{1}{\sqrt{2}} \left( \hat{b}(h_0^{-1/2}f) + \hat{a}^{\dagger}(h_0^{-1/2} f)\right) \hspace{1cm}
\hat{\pi}^{\dagger}(f) := \frac{1}{\sqrt{2}i} \left( \hat{b}(h_0^{1/2}f) - \hat{b}^{\dagger}(h_0^{1/2} f)\right).
\end{eqnarray}
where $h_0 = \sqrt{-\partial_x^2 + m^2}$ is the Hamiltonian operator for a free relativistic particle, defined on the Hilbert space $H_1 = \bar{H}_1$.

We construct the   Hamiltonian operator $\hat{H}$ for the field by substituting the field operators above into the classical Hamiltonian density (\ref{hamden}). After normal ordering,
\begin{eqnarray}
\hat{H} = \int dx dx' \left(\hat{a}^{\dagger}(x)h_1(x,x') \hat{a}(x) + \hat{b}^{\dagger}(x)h_2(x,x') \hat{b}(x)\right). \label{hamilton}
\end{eqnarray}
In Eq. (\ref{hamilton}),  $h_{1}(x,x')$ are matrix elements in the position basis of a Hamiltonian operator $h_1$ on the  Hilbert space $H_1$ of a single particle and $h_2(x,x')$ are matrix elements in the position basis of a Hamiltonian operator $h_2$ on the Hilbert space $\bar{H}_1$ of a single antiparticle. The operators $h_{1,2}$ are defined as
\begin{eqnarray}
h_{1} = h_0 + \tilde{V}, \hspace{1cm}
h_2 = h_0 - \tilde{V} \label{h1h2}
\end{eqnarray}
where
\begin{eqnarray}
\tilde{V} := \frac{1}{2} \left( h_0^{1/2} V h_0^{-1/2} + h_0^{1/2} V h_0^{-1/2}\right) = V + [[V,h_0^{1/2}], h_0^{-1/2}], \label{tildev}
\end{eqnarray}
 is the  potential operator, that includes a QFT correction term.  The sign difference between $h_1$ and $h_2$ indicates the fact that a potential barrier for particles is a potential well for anti-particles.

In what follows, we will consider a positive-valued potential $V(x)$, such that the corresponding single-particle Hamiltonian $h_1$ has purely continuous spectrum. The eigenstates of $h_1$  are characterized by energies $E > m$ and they have a double degeneracy. We denote the pair of eigenstates corresponding to the same value $E$ of energy as $f_{k+}$ and $f_{k-}$, where $ k = \sqrt{E^2- m^2} > 0$. The eigenstates $f_{k+}(x)$ correspond to only positive momentum flux at $x = \infty$ and they are of the form
\begin{eqnarray}
f_{k+}(x) = \frac{1}{\sqrt{2\pi}}  \left\{ \begin{array}{cc} e^{ikx} + R_k e^{-ikx} & x < -\frac{d}{2} \\
T_k e^{ikx} & x > \frac{d}{2}  \end{array} \right. \label{f+}
\end{eqnarray}
 In Eq. (\ref{f+}), $T_k$ and $R_k$ are the usual transmission and reflection coefficients for a right-moving plane wave of momentum k. The eigenstates $f_{k-}$ are orthogonal to $f_{k+}$ and they satisfy
\begin{eqnarray}
f_{k-}(x) = \frac{1}{\sqrt{1 - w_k^2}} \left[ - w_k f_{k+}(x) + f_{k+}(-x) \right],
\end{eqnarray}
where $w_k = \frac{1}{2} (T^*_kR_k + T_kR^*_k)$ is essentially  the overlap between $f_{k+}$ and its parity-transform. For parity-symmetric potentials, i.e., for potentials such that $V(x) = V(-x)$, the coefficient $w_k$ vanishes and $f_{k-}(x) = f_{k+}(-x)$.

 The Hamiltonian $h_2$, describing anti-particles, has both continuous  ($E > m$) and discrete spectrum ($E < m$). We denote the continuous-spectrum eigenstates as $g_{k+}$ and $g_{k-}$. They are similar in structure to the generalized eigenstates $f_{k \pm}$ of $h_1$. Their form is not relevant to the purposes of this paper.

  We denote the discrete-spectrum eigenstates of $h_2$ as $g_n(x)$, where the label $n$ runs into a finite set of values.
 The eigenstates $g_n(x)$ decay exponentially outside the barrier region.
\begin{eqnarray}
g_n(x) = \left\{ \begin{array}{cc} A_n e^{\gamma_n x} & x < -d/2  \\
B_n e^{-\gamma_n x} & x > d/2 \end{array}  \right.,
\end{eqnarray}
for some positive constants $A_n$ and $B_n$, and
\begin{eqnarray}
\gamma_n = \sqrt{m^2 - E_n^2}. \label{gaman}
 \end{eqnarray}
 For parity-symmetric potentials, $B_n = P(n) A_n$, where $P(n)$ is the parity of the eigenstate $g_n(x)$.

Next, we define the energy-basis creation and annihilation operators,
\begin{eqnarray}
\hat{a}_{k\pm} &=& \int dx \hat{a}(x) f^*_{k\pm}(x) \hspace{1cm} \hat{b}_{k \pm} =  \int dx \hat{b}(x) g^*_{k\pm}(x) \hspace{1cm} \hat{b}_n = \int dx \hat{b}(x) g^*_{n}(x)\\
\hat{a}^{\dagger}_{k\pm} &=& \int dx \hat{a}^{\dagger}(x) f_{k\pm}(x) \hspace{0.85cm} \hat{b}^{\dagger}_{k \pm} =  \int dx \hat{b}^{\dagger}(x) g_{k\pm}(x) \hspace{0.85cm} \hat{b}^{\dagger}_n = \int dx \hat{b}^{\dagger}(x) g_{n}(x).
 \end{eqnarray}
 Then, the Heisenberg-picture field operators are
\begin{eqnarray}
\hat{\phi}(x, t) = \sum_{\sigma = -}^+ \int_0^{\infty} \frac{dk}{\sqrt{2E_k}} \left(\hat{a}_{k,\sigma} f_{k, \sigma}(x) e^{-i E_k t} + \hat{b}^{\dagger}_{k,\sigma} g^*_{k, \sigma}(x) e^{i E_kt} \right) + \sum_n \frac{1}{\sqrt{2E_n}} \hat{b}^{\dagger}_n g_n^*(x) e^{iE_nt} \label{fi1} \\
\hat{\phi}^{\dagger}(x,t) = \sum_{\sigma = -}^+ \int_0^{\infty} \frac{dk}{\sqrt{2E_k}} \left(\hat{b}_{k,\sigma} g_{k, \sigma}(x) e^{-i E_k t} + \hat{a}^{\dagger}_{k,\sigma} f^*_{k, \sigma}(x) e^{i E_kt} \right) + \sum_n \frac{1}{\sqrt{2E_n}} \hat{b}_n g_n(x) e^{-iE_nt}. \label{fi2}
\end{eqnarray}

Eqs. (\ref{fi1}) and (\ref{fi2}) are the basis for the QFT treatment of tunneling time in the following section.

\section{Probabilities for tunneling time}
In this section, we derive a general expression for the detection probability $P(L,t)$ for relativistic particles tunneling through a potential barrier. First, we present a brief review of the QTP method, and then we derive an equation that relates the detection probability to the correlation functions of a quantum field theory. Finally, we specialize to the scalar field system described in Sec. 2, and we  express the probability $P(L,t )$ in terms of the transmission and reflection coefficient associated with the potential barrier.

\subsection{Background}

Our approach is based on the general methodology for constructing time-of-arrival probabilities associated with general detectors, that was developed in Ref. \cite{AnSav12} (the Quantum Temporal Probabilities method). The reader is referred to this article for a detailed presentation of the method. Here, we present a brief review, setting up the background for its application to the tunneling-time problem in Secs. 3.2 and 3.3.

The key result of Ref. \cite{AnSav12} is the derivation of a general formula for probabilities associated with the time of an event  in a general quantum system. When we use the word "event", we mean the appearance of a definite and persistent macroscopic record of observation, essentially the irreversible amplification of a microscopic process that corresponds to a quantum measurement. The event time $t$ is a coarse-grained, quasi-classical parameter associated with such records: it corresponds to the reading of an external classical clock simultaneous with the emergence of the record.

Let  ${\cal H}$ be the
Hilbert associated with  the physical   system under consideration. We incorporate  the measurement device into  the quantum description, so  ${\cal H}$ describes the degrees of freedom of both the microscopic particles and the macroscopic measurement apparatus.
 We assume that
 ${\cal H}$ splits into two subspaces: ${\cal H} = {\cal
H}_+ \oplus {\cal H}_-$. The subspace ${\cal H}_+$ describes the accessible
states of the system given that a specific event is realized; the subspace ${\cal H}_-$ is the complement of ${\cal H}_+$. For example, if the quantum event under consideration is a detection of a particle by a macroscopic apparatus, the subspace ${\cal H}_+$ corresponds to all accessible states of the apparatus given that a detection event has been recorded. We denote  the
projection operator onto ${\cal H}_+$ as $\hat{P}$ and
the projector onto ${\cal H}_-$ as $\hat{Q} := 1  - \hat{P}$.

 We note that the transitions under consideration are always correlated with the emergence of a macroscopic observable (including time) that is recorded as a measurement outcome.  In modeling quantum measurements, the relevant transitions are associated to the degrees of freedom of the measurement apparatus, and not of the microscopic system. In this sense, the transitions considered here are {\em irreversible}. Once they occur, and a measurement outcome has been recorded, the further time evolution of the degrees of freedom in the measurement device is irrelevant to the probability of transition.

Once a transition has taken place,   the values of a microscopic variable may be determined through correlations with a pointer variable of the measurement apparatus. We denote by $\hat{P}_\lambda$   projection operators (or, more generally, positive operators) corresponding to different values $\lambda$ of a physical magnitude that can be measured only if the quantum event under consideration has occurred. For example, when considering transitions associated with particle detection, the projectors $\hat{P}_\lambda$  may be correlated  to properties of the microscopic particle, such as position, momentum and energy.
The set of projectors $\hat{P}_\lambda$ is exclusive ($\hat{P}_{\lambda} \hat{P}_{\lambda'} = 0, $ if $\lambda \neq \lambda'$). It is also exhaustive given that the event under consideration  has occurred; i.e., $\sum_\lambda \hat{P}_\lambda = \hat{P}$.

We also assume that  the system is initially ($t = 0$) prepared at a
state $|\psi_0 \rangle \in {\cal H}_-$, and that the time evolution is
governed by the self-adjoint Hamiltonian operator $\hat{H}$.

In Ref. \cite{AnSav12}, we derived
 the probability amplitude $| \psi; \lambda, [t_1, t_2] \rangle$ that, given an initial state $|\psi_0\rangle$, a transition occurs at some instant in the time interval $[t_1, t_2]$ and a recorded value $\lambda$ is obtained for some observable:

 \begin{eqnarray}
| \psi_0; \lambda, [t_1, t_2] \rangle = - i e^{- i \hat{H}T}
\int_{t_1}^{t_2} d t \hat{C}(\lambda, t) |\psi_0 \rangle.
\label{ampl5}
\end{eqnarray}

where   the {\em class operator} $\hat{C}(\lambda, t)$ is defined as
\begin{eqnarray}
\hat{C}(\lambda, t) = e^{i \hat{H}t} \hat{P}_{\lambda} \hat{H}
\hat{S}_t, \label{class}
\end{eqnarray}
and
\begin{eqnarray}
\hat{S}_t =  \lim_{N \rightarrow \infty}
(\hat{Q}e^{-i\hat{H} t/N} \hat{Q})^N \label{restricted}
\end{eqnarray}
is the restriction of the propagator into ${\cal H}_-$. The parameter $T$ in Eq. (\ref{ampl5}) is a reference time-scale at which the amplitude is evaluated. It defines an upper limit to $t$ and it corresponds to the duration of an experiment. It cancels out when evaluating probabilities, so it does not appear in the physical predictions.

If $[\hat{P}, \hat{H}] = 0$, i.e., if the Hamiltonian
evolution preserves the subspaces ${\cal H}_{\pm}$, then $|\psi_0;
\lambda, t \rangle = 0$. For a Hamiltonian    of the form $\hat{H} = \hat{H_0} + \hat{H_I}$, where $[\hat{H}_0, \hat{P}] = 0$, and $H_I$ a perturbing interaction, we obtain, to leading order in the perturbation
\begin{eqnarray}
\hat{C}(\lambda, t) = e^{i \hat{H}_0t} \hat{P}_{\lambda} \hat{H}_I
e^{-i \hat{H}_0t}. \label{perturbed}
\end{eqnarray}
The benefit of Eq. (\ref{perturbed}) is that it does not involve the restricted propagator $\hat{S}_t$, which is difficult to compute.

The amplitude  Eq. (\ref{ampl5}) squared defines  the probability $p (\lambda, [t_1, t_2])\/$
that at some time in the interval $[t_1, t_2]$ a detection with
outcome $\lambda$ occurred
\begin{eqnarray}
P(\lambda, [t_1, t_2]) := \langle \psi_0; \lambda, [t_1, t_2] | \psi_0;
\lambda, [t_1, t_2] \rangle =   \int_{t_1}^{t_2} \,  dt
\, \int_{t_1}^{t_2} dt' \; Tr [\hat{C}(\lambda, t) \hat{\rho}_0
\hat{C}^{\dagger}(\lambda, t) ], \label{prob1}
\end{eqnarray}
where $\hat{\rho}_0 = |\psi_0\rangle \langle \psi_0|$.

However, the expression $P(\lambda, [t_1, t_2])$ does not correspond to a well-defined probability measure, because it
fails to satisfy the Kolmogorov additivity condition for probability measures
\begin{eqnarray}
P(\lambda, [t_1, t_3]) = P(\lambda, [t_1, t_2]) + P(\lambda, [t_2,
t_3]).  \label{kolmogorov}
\end{eqnarray}

Eq. (\ref{kolmogorov}) does not hold for generic choices of $t_1, t_2$ and $t_3$.   However, in a macroscopic system (or in a system with a macroscopic component) one expects that Eq. (\ref{kolmogorov}) holds with a good degree of approximation, given a sufficient degree of coarse-graining \cite{GeHa93, hartlelo}. Thus, if the time of transition is associated with a macroscopic measurement record, there exists a coarse-graining time-scale $\sigma$, such that    Eq. (\ref{kolmogorov}) holds, for $ |t_2 - t_1| >> \sigma$ and $|t_3 - t_2| >> \sigma$. Then, Eq. (\ref{prob1}) does define a probability measure when restricted to intervals of size  larger than $\sigma$.

Eq. (\ref{prob1}) simplifies if there is a separation of time scales between the macroscopic scale of observation, the coarse-graining time scale and the characteristic timescales of the microscopic system. In this case, it can be shown \cite{AnSav12} that the probabilities Eq. (\ref{prob1})  are obtained  from the probability density

\begin{eqnarray}
P(\lambda, t) = \int d\tau  Tr\left[\hat{C}(\lambda, t+\frac{\tau}{2})
 \hat{\rho}_0 \hat{C}^{\dagger}(\lambda, t- \frac{\tau}{2})\right] \label{pp2}
\end{eqnarray}

The probability density Eq. (\ref{pp2}) is of the form
 $Tr[\hat{\rho}_0 \hat{\Pi}(\lambda, t)]$,
where
\begin{eqnarray}
\hat{\Pi}(\lambda, t) = \int  d\tau \hat{C}^{\dagger}(\lambda, t - \frac{\tau}{2})
\hat{C}(\lambda, t + \frac{\tau}{2}). \label{povm2}
\end{eqnarray}

The positive
operator
\begin{eqnarray}
\hat{\Pi}_{\tau}(N) = 1 - \int_0^{\infty} dt \int d \lambda
\hat{\Pi}_{\tau}(\lambda, t), \label{nodet}
\end{eqnarray}
 corresponds to the alternative $ N$ that no detection took place
in the time interval $[0, \infty)$. The positive operator $\hat{\Pi}_{\tau}(N)$ together with the positive operators Eq. (\ref{povm2})
 define a Positive-Operator-Valued Measure (POVM). The POVM Eq. (\ref{povm2}) determines the probability density that a transition took place at time $t$, and that the outcome
$\lambda$ for the value of an observable has been recorded.

\subsection{Temporal probabilities from field correlation functions}

Next, we employ Eq. (\ref{pp2}) for constructing probabilities associated with time-of-arrival measurements in a quantum field theoretic set-up.  The system under consideration consists  of a quantum field $\hat{\phi}$ that describes microscopic particles  and a macroscopic particle detector. Eventually, we will identify the quantum field with the scalar field $\hat{\phi}(x)$, Eq. (\ref{fi1}), associated with tunneling, but the results obtained in Sec. 3.2 apply for any quantum field theory. With small modifications, the method applies also to spinor and vector fields. The derivation presented here is an adaptation of the method described in Ref. \cite{AnSav12}.

The Hilbert space for the total system is the tensor product ${\cal F} \otimes {\cal H}_a$. The Hilbert space ${\cal H}_a$  describes the apparatus's degrees of freedom and ${\cal F}$  is the Hilbert space associated with the field $\hat{\phi}$.

The Hamiltonian of the total system  is the sum $\hat{H}_0 \otimes 1 + 1 \otimes \hat{H}_{a} + \hat{H}_{I}$. $\hat{H}_0$ is the Hamiltonian  that describes the quantum field $\hat{\phi}$, $\hat{H}_{a}$ describes the dynamics of the apparatus and $\hat{H}_{I}$ is an interaction term.

The requirements of Lorentz covariance and unitarity in quantum field theory imply that the interaction Hamiltonian $\hat{H}_{I}$ must be a local functional of the fields $\hat{\phi}(x)$ and $\hat{\phi}^{\dagger}(x)$. Therefore, we consider an interaction Hamiltonian of the form
\begin{eqnarray}
\hat{H}_{I} = \int dx \left[\hat{\phi}(x) \otimes \hat{J}^{\dagger}(x) + \hat{\phi}^{\dagger}(x) \otimes \hat{J}(x) \right], \label{hint}
\end{eqnarray}
where $\hat{J}$ and $\hat{J}^{\dagger}$ are current operators on the Hilbert ${\cal H}_a$ of the detector. The interaction Hamiltonian Eq. (\ref{hint}) corresponds to particle detection by absorption.

We also assume  an initial state $|\Psi_0\rangle $ of the detector such that
\begin{eqnarray}
\hat{J}(x) |\Psi_0\rangle = 0. \label{jps}
 \end{eqnarray}
 This condition guarantees that the detector is sensitive to particles rather than anti-particles. To see this, note that the term $\hat{\phi}^{\dagger}(x) \otimes \hat{J}(x) $ in the Hamiltonian Eq. (\ref{hint}) corresponds to the processes of either a particle being created at the measurement device or an anti-particle being annihilated. The second process is negligible if the initial state contains only particles at energies $E < m$. The first process is not desirable in a particle detector. Eq. (\ref{jps}) guarantees that the only transitions in the detector are caused by the interaction with microscopic particles.

  Next, we  specify the macroscopic variables associated with particle detection. These correspond to degrees of freedom of the macroscopic apparatus and they are expressed in terms of the
 positive operators $1 \otimes \hat{\Pi}_{L}$ on ${\cal F}(H_1\oplus \bar{H}_1) \otimes {\cal H}_a$, labeled by the values $L$ of a macroscopic observable correlated to a particle's position in the laboratory frame of reference.
 The operators  $\hat{\Pi}_{L}$ are defined on
 ${\cal H}_{a}$ and they satisfy the completeness relation $\int d L \hat{\Pi}_{L} = \hat{P}$, where $\hat{P}$  is the projector onto the subspace ${\cal H}_+$. We choose  $\hat{\Pi}_{L}$ so that $\sqrt{\hat{\Pi}_L}$  corresponds to an unsharp Gaussian sampling of position at $X = L$,

\begin{eqnarray}
\sqrt{\hat{\Pi}_L} = \frac{1}{(\pi \delta^2)^{1/4}}\sum_a \int dX e^{-\frac{(L-X)^2}{2 \delta^2}} |X, a\rangle \langle X,a |, \label{po}
\end{eqnarray}

The index $a$ in Eq. (\ref{po}) refers to the degrees of freedom of the apparatus other than the pointer variable, and $\delta$ corresponds to the spatial resolution of the detector.

 We place no restriction on the apparatus' Hamiltonian $\hat{H}_a$, except for the requirement that it commutes with the operator $\hat{P}$: $[\hat{H}_{a}, \hat{P} ]   = 0$.  This implies that the self-evolution of the apparatus degrees of freedom does not lead to spurious detection records.
 It follows that $[1\otimes \hat{P}, \hat{H}_s \otimes 1 + 1 \otimes \hat{H}_{a}] = 0$; hence, the class operators $\hat{C}(\lambda, t)$ follow from  Eq. (\ref{perturbed}).

Substituting into Eq. (\ref{povm2}), we obtain
\begin{eqnarray}
P(L,t) = \int d\tau  \int dx dx' \langle \psi_0| \hat{\phi}^{\dagger}(x', t-\tau/2) \hat{\phi}(x, t + \tau/2) | \psi_0 \rangle R(L, \tau, x, x') \label{plt01}
\end{eqnarray}
where $\hat{\phi}(x,t)$ and $\hat{\phi}^{\dagger}(x,t)$ are the Heisenberg-picture fields,
$|\psi_0\rangle$ is the initial field state and
\begin{eqnarray}
R(L, \tau, x, x') = \langle \Psi_0| \hat{J}( x') \sqrt{\hat{\Pi}_{L}} e^{ i \hat{H}_{a}\tau} \sqrt{\hat{\Pi}_{L}} \hat{J}^{\dagger}( x)|\Psi_0\rangle. \label{rkernel}
\end{eqnarray}
is a kernel that involves the detailed physics of the detector.

 The macroscopic record of a detection must be correlated with the locus of the interaction, in the sense that $\sqrt{\hat{\Pi}_{L}} \hat{J}^{\dagger}( x)|\Psi_0\rangle \simeq 0$ if $|L - x|$ is much larger than the detector's resolution $\delta$. This implies that $x$ and $x'$ in Eq. (\ref{rkernel}) lie within a distance $\delta$ from $L$, otherwise the kernel $R(L, \tau, x, x')$ vanishes. Equivalently, the kernel $R$ vanishes unless $X = (x+x')/2$ lies within distance of order $\delta $ from $L$, and $|z|$ is of order $\delta$.
 If $\delta$ is much smaller than the distance between source and detector, we can approximately  set $L = (x+x')/2$, so that
\begin{eqnarray}
R(L, \tau, x, x') \simeq \delta (L - \frac{x+x'}{2}) g( x-x', \tau),
\end{eqnarray}
This form for $R(L,\tau, x, x')$ is consistent  with several different detector models that have been constructed in Ref. \cite{AnSav12}.
The kernel $g(z, \tau)$ depends on the detailed physics of the detector. It vanishes for values of  $|z|$ larger than the position resolution $\delta$ of the detector and  $|\tau|$ larger than the decoherence time of the detector, respectively.

Eq. (\ref{plt01}) becomes
\begin{eqnarray}
P(L,t) = \int d\tau \int dz \langle \psi_0|\hat{\phi}^{\dagger}(L -\frac{z}{2}, t-\frac{\tau}{2}) \hat{\phi}(L+\frac{z}{2}, t+\frac{\tau}{2})|\psi_0\rangle g(z, \tau). \label{prob1b}
\end{eqnarray}

It is important to observe that no assumption about the form of the Hamiltonian  $\hat{H}_0$, or about the initial state $| \psi_0 \rangle $ has been made for the derivation of Eq. (\ref{plt01}) or Eq. (\ref{prob1b}). Eq. (\ref{plt01}) reveals a relation between detection probability and field correlation functions that applies to any quantum field theory. The probability of detection is determined by the two-point function of the fields that interact with the detectors.

\subsection{The probability distribution for the time of detection}
Next, we specialize to the case of the fields $\hat{\phi}(x,t)$, $\hat{\phi}^{\dagger}(x,t)$ of Eqs. (\ref{fi1}---\ref{fi2}), associated with a tunneling set-up. We consider a single-particle initial state for the field $|\psi_0 \rangle = \int dx \hat{a}^{\dagger}(x) \psi_0(x) |0\rangle$, where $|0 \rangle $ is the field vacuum and $\psi_0(x)$ a single-particle wave function concentrated on positive values of momentum.

Substituting Eqs. (\ref{fi1}---\ref{fi2})  into Eq. (\ref{prob1b}), we obtain
\begin{eqnarray}
P(L, t) = P_0(L, t) + P_1(t) + P_2(L, t),
\end{eqnarray}

 where
\begin{eqnarray}
P_0(L, t) &=& \int d \tau \int dz g(z, \tau) \left[ \int dx \Delta(L+\frac{z}{2},x; \tau + \frac{\tau}{2} ) \psi_0(x)\right] \nonumber \\
&\times& \left[ \int dx' \Delta(L-\frac{z}{2},x'; \tau - \frac{\tau}{2} ) \psi_0(x')\right]^* \\
P_1(L, t) &=& \int d \tau \int dz g(z, \tau)\sum_{\sigma} \left(\int_{-\infty}^{\infty} \frac{dk}{(2\pi)(2E_k)} e^{-ikz + iE_k \tau}\right)\\
P_2(L, t) &=& \int d \tau \int dz g(z, \tau)D(L -\frac{z}{2}, L+ \frac{z}{2}; -\tau),
\end{eqnarray}
and
\begin{eqnarray}
\Delta(x,x';t) &=& \sum_{\sigma = -}^+ \int_0^{\infty} \frac{dk}{\sqrt{2 E_k}}   f_{k_\sigma}(x) f^*_{k_\sigma}(x')e^{-iE_kt} \\ \label{deltaxx}
D(x, x' ; t) &=& \sum_n \frac{1}{2E_n} g_n^*(x) g_n(x') e^{-iE_n t}. \label{dxx2}
\end{eqnarray}

The  terms $P_1(L, t)$ and $P_2(L, t)$   are independent of the initial state of the particle.They correspond to `false alarms', that is, spurious detection events---such terms are generic in the theory of relativistic quantum measurements \cite{PerTer}. They act as noise that obscures the signal term $P_0(L, t)$.     The $P_1$ term in is independent on the location of the detector. It is also independent of the barrier, as it persists for $V(x) = 0$. Its contribution to the total probability is negligible for detector sizes much larger than the de-Broglie wave-length of particles. The  term $P_2$ originates from the anti-particle bound states and it depends on the position $L$ of the detector. However, the function $D_2(x,x',t)$, Eq. (\ref{dxx2})  decays exponentially outside the barrier region, with a rate given by $\gamma_n$, Eq. (\ref{gaman}). Therefore, the term $P_2(L, t)$ is strongly suppressed for $L >> \max_n\{\gamma_n^{-1}\}$. The largest values of $\gamma_n$ are of the order of the length $\sqrt{d/m} << d$.

 For a detector far from the barrier ($L >> d$), the contributions from the false-alarm terms $P_1$ and $P_2$ becomes negligible, and  $P(L, t) = P_0(L, t)$.

The term $P_0(L, t)$ involves the function $\Delta(x, x'; t)$, Eq. (\ref{deltaxx}).
For $x > \frac{d}{2}$, $x' < -\frac{d}{2}$, we obtain
\begin{eqnarray}
\Delta(x, x'; t) = \int_0^{\infty} \frac{dk }{(2 \pi)\sqrt{2E_k}} \left[ e^{i k(x-x')} (T_k + w_k \frac{- R_k +w_k T_k}{1 - |w_k|^2}) + e^{-ik(x+x')} \frac{-w_k}{1 - w_k^2}  \right. \nonumber \\
\left. + e^{ik(x+x')} \frac{R_k^* +(T_k^* -w_k R_k^*)(R_k -w_kT_k)}{1 - w_k^2} + e^{-ik(x-x')} \frac{T^*_k-w_k R^*_k}{1 - |w_k|^2} \right] e^{-iE_kt}. \label{delta2}
\end{eqnarray}
The term in the second line of Eq.  (\ref{delta2})  vanishes when $\Delta(x, x'; t)$ acts on wave functions with support on positive values of momentum. The second term in the first line of Eq.  (\ref{delta2}) is strongly suppressed for $x \simeq L$ and $t > 0$.

Thus, for sufficiently large $L$, the probability density $P(L,t)$ becomes
\begin{eqnarray}
P(L, t) = \int_0^{\infty}  \frac{dk}{(2\pi)\sqrt{2E_k}} \int_0^{\infty} \frac{dk'}{(2\pi)\sqrt{2E_{k'}}} f(k,k') A_k A^*_{k'} \tilde{\psi}_0(k) \tilde{\psi}_0(k') e^{i(k-k')L-i(E_k-E_{k'})t }, \label{prob3}
\end{eqnarray}
where
\begin{eqnarray}
A_k = \frac{T_k -w_k R_k}{1 - w_k^2} , \label{ak}
\end{eqnarray}
\begin{eqnarray}
f(k, k') = \int d \tau \int dz g(\tau, z) e^{i \frac{k+k'}{2} z + i \frac{E_k+E_{k'}}{2}\tau},
\end{eqnarray}
$\tilde{\psi}_0(k)$ is the Fourier transform of the initial state, and $E_k = \sqrt{k^2+m^2}$.

For an initial state with support on positive momenta, $P(L, t)$ is strongly suppressed for $t < 0$. Hence, we approximate $\int_0^{\infty}P(L,t) \simeq \int_{-\infty}^{\infty}P(L,t)$. For a free particle ($A_k  = 1$),  the time-integrated probability for a particle to be detected at $x =L$ is
\begin{eqnarray}
\int_0^{\infty}P(L,t)  = \int_0^{\infty} \frac{dk}{2\pi} |\tilde{\psi}_0(k)|^2 \frac{f(k,k)}{2E_k v_k}.
\end{eqnarray}
The quantity $\alpha(k) = \frac{f(k,k)}{2E_k v_k}$ then defines the absorption coefficient of the detector for particles with momentum $k$, i.e., the fraction of the number of incident particles absorbed by the detector. The function $f(k,k')$ depends on $k$ and $k'$ only through the combinations $(k+k')/2$ and $(E_k +E_{k'})/2$. For an initial state with momentum spread $\delta k$ much smaller than the mean momentum $\bar{k}$, we approximate
\begin{eqnarray}
f(k,k') = \sqrt{|f(k,k)|} \sqrt{|f(k',k')|}.
\end{eqnarray}

 Then, Eq. (\ref{prob3}) becomes
 \begin{eqnarray}
P(L,t) = \left| {\cal A}(L, t) \right|^2.
\label{plt}
\end{eqnarray}

where
\begin{eqnarray}
{\cal A}(L,t) = \int_0^{\infty} \frac{dk}{2\pi} \sqrt{\alpha(k) |v_k|} A_k \tilde{\psi}_0(k) e^{ikL-i E_k t} \label{amplmain}
\end{eqnarray}

Eqs. (\ref{plt}---\ref{amplmain}) are the main result of this paper. They express the probability distribution for particle detection in a tunneling experiments in terms of the transmission and reflection coefficients of the barrier, encoded in the coefficient $A_k$. All information about the detector is encoded in the  absorption coefficient $\alpha(k)$ for particles of momentum $k$ and all information about the barrier in the complex amplitude $A_k$.

In absence of the potential barrier $A_k = 1$, and Eq. (\ref{plt}) reduces to the time-of-arrival distribution for free relativistic particles derived in Ref. \cite{AnSav12}, which generalizes Kijowki's probability distribution for the time of arrival of non-relativistic particles \cite{Kij}.

The time-integrated probability Eq. (\ref{plt}) is
\begin{eqnarray}
\int_{-\infty}^{\infty} dt P(L,t) = \int \frac{dk}{2\pi} \alpha(k) |A_k|^2 |\tilde{\psi}_0(k)|^2. \label{tint}
\end{eqnarray}
Eq. (\ref{tint}) implies that $|A_k|^2$ is the transmission probability, i.e., the probability that a particle of momentum $k$ crosses the barrier at any time $t$. For parity symmetric potentials, $A_k = T_k$ and the transmission probability coincides with the standard expression $|T_k|^2$. In general, however, they differ. The reason is that in the QFT description of tunneling, both right-moving and left-moving modes contribute to field correlation functions, and by virtue of Eq. (\ref{prob1b}), they contribute to the detection probability.

 It is convenient to parameterize $T_k = |T_k| e^{i \phi_k}$ and $R_k = - i |R_k| e^{i \phi_k + i \chi_k}$. In parity-symmetric potentials, $\chi_k = 0$. In the regime of opaque barrier, $|T_k| << |R_k| \simeq 1$, hence, to leading order in $|T_k|$
\begin{eqnarray}
A_k \simeq \frac{1}{2} |T_k| (1 + e^{2 i \chi_k}).  \label{akappr}
\end{eqnarray}
We then obtain
\begin{eqnarray}
|A_k|^2 \simeq |T_k|^2 \cos^2\chi_k
\end{eqnarray}
For potentials characterized by large deviations from parity symmetry, the transmission probability $|A_k|^2$ may differ significantly from  $|T_k|^2$. In fact, for $\chi_k = \frac{\pi}{2}$, the transmission probability vanishes even if $|T_K|^2$ is non-zero.

\subsection{Delay time}

Next, we examine the case of an initial state localized around $x = - x_0$ and concentrated around the value $k = p $ for momentum. In particular, we consider a wave function $\tilde{\psi}_0(k) = \tilde{u}_0(k - p) e^{ikx_0}$, where $\tilde{u}_0(k)$ is a positive-valued, square-integrable function centered around $k = 0$. Then, the integrand in Eq. (\ref{plt}) involves a rapidly oscillating phase $\exp [ i k (x_0 + L) - i E_k t + \theta_k]$, where $\theta_k = Im \log A_k$. Often, but not always, this implies that the integral is strongly suppressed unless the phase is stationary for $k  = p$, i.e., unless
\begin{eqnarray}
x_0 + L - v_p t + \theta'_p = 0. \label{stationary}
\end{eqnarray}
  In general, the stationary phase approximation is valid   if the probability density  Eq. (\ref{plt}) has a single peak. In this case, the solution to Eq.   (\ref{stationary}), $\bar{t}(p) := (x_0 +L + \theta'_p)/v_p$, determines the  location of the peak.

  In absence of the barrier, and for the same initial state, the time-of-arrival probability distribution is peaked around $\bar{t}_0(p) = (x_0 +L)/v_p$. Thus, when comparing two time-of-arrival experiments, one with the barrier and one without,  particles in the former are detected with a delay
\begin{eqnarray}
t_d(p) = \bar{t} - \bar{t}_0(p) = \theta'_p/v_p = \frac{1}{v_k} Im \left[\frac{\partial  \log A_k}{\partial k} \right]_{k=p}. \label{delt}
\end{eqnarray}
 when compared with detected particles in the latter.
For parity symmetric potentials, $A_k = T_k$ and the delay time Eq. (\ref{delt})  coincides with the Bohm-Wigner phase time \cite{BW}. In general, they differ.

Given Eq. (\ref{delt}) one may infer that the time particles spent inside the barrier region equals
\begin{eqnarray}
\tau_p =  t_d(p) - d/v_p. \label{tunt}
 \end{eqnarray}

 However, this inference is based on an analogy from classical physics and it does not follow from the rules of quantum theory. The time delay $t_d(p)$, Eq. (\ref{delt}), is obtained from the comparison of probability distributions from different experiments; it is not directly measured in a single experiment. The delay time is not a quantum observable or a random variable of the theory. It  is a temporal parameter that appears in the probability distribution $P(L,t)$ for a specific class of initial states. This parameter may incorporate significant information about  properties of the barrier, but only if it captures a significant physical feature of the probability density $P(L,t)$.

Moreover, Eq. (\ref{delt}) follows from a saddle-point approximation to the probability density $P(L,t)$, Eq. (\ref{plt}) that is valid only if the integrand in Eq. (\ref{plt})  has a single maximum. This is not true in general, as we will see in the study of the double square barrier in Sec. 4.3. In such cases, the naive application of the saddle point approximation leads to erroneous physical conclusions. In general, higher moments of the probability distribution $P(L,t)$ contain significant information and the temporal aspects of the tunneling process cannot be adequately described in terms of a single parameter such as the delay time.

In the following section, we will refer to $\tau_p$, Eq. (\ref{tunt}), as "tunneling time", for convenience, without committing  to  its interpretation as the transit time of the particle through the barrier region. We will view $\tau_p$  as a useful time parameter that characterizes the probability distribution $P(L,t)$. We will take up the physical interpretation of $\tau_p$ in Sec. 5, where we will present our proposed resolution to the superluminality paradox in tunneling.

\section{Application: Piecewise constant potentials }
In this section, we construct the probability distribution for the time of detection for two important examples of tunneling potentials: the square barrier and the symmetric double square barrier. Both potentials are parity-symmetric, so the amplitude $A_k$, Eq. (\ref{ak}), coincides with the transmission amplitude $T_k$.

  We  consider an initial state $\tilde{\psi}_0(k)$, centered around $x = -x_0$ and well concentrated around a mean momentum $k = p$,
  \begin{eqnarray}
  \tilde{\psi}_0(k) = \tilde{u}_0(k-p) e^{ikx_0},
   \end{eqnarray}
   where $\tilde{u}_0$ is an even, positive-valued wave function centered around $k = 0$ with width $\sigma_p$. We assume that the absorption coefficient $\alpha(k)$ does not vary significantly with momentum, so that we can set it equal to unity. Then, Eq. (\ref{plt}) takes the form
  \begin{eqnarray}
  P(L,t) = \left| \int_0^{\infty} \frac{dk}{2\pi} \sqrt{ |v_k|} T_k \tilde{u}_0(k-p) e^{ik(L+x_0) -i E_k t}\right|^2. \label{parplt}
\end{eqnarray}

In what follows, we evaluate Eq. (\ref{parplt}) for the square barrier and the symmetric double square barrier.

\subsection{Junction conditions}
Explicit calculations of tunneling time are usually performed for piecewise constant potentials. In such potentials, eigenstates of the Hamiltonian are identified from junction conditions on the points of discontinuity. The Hamiltonians $h_1$ and $h_2$, Eq. (\ref{h1h2})  are non-local operators, and their junction conditions turn out to be different from the ones usually considered in the literature. Here, we present the derivation of the junction conditions.

 We consider the Hamiltonian $h_1$, with a piecewise constant potential. For piecewise constant potentials, $\tilde{V}(x) = V(x)$   except for the points of discontinuity.  Hence, except for the points of discontinuity, the eigenfunctions of $h_1$  are of the form $A e^{ikx} + B e^{-ikx} $ where $k$ may be either real or imaginary. Let us assume that $x = 0$ is a point of discontinuity, and consider a neighborhood $U$ around $x = 0$ with no other point of discontinuity. Let $g(x)$ be an eigenstate of $h_1$ with energy $E$. Assuming $x \in U$, $g''(x) = -k_1^2 g(x)$ for $x < 0$, and $g''(x) = - k_2^2 g(x)$ for $x >0$, where $k_1$ and $k_2$ are real or imaginary constants. Then,
\begin{eqnarray}
\int_{-\epsilon}^{\epsilon}dx h_1\psi(x) = E \int_{-\epsilon}^{\epsilon}dx \psi(x),
\end{eqnarray}
for some constant $\epsilon$ that will be eventually taken to zero. For continuous $g(x)$, all terms in the above equation that do not involve derivatives vanish at the limit $\epsilon \rightarrow 0$. Hence, as $\epsilon \rightarrow 0$,
\begin{eqnarray}
\int_{-\epsilon}^{\epsilon}dx \sqrt{-\partial_x^2+m^2}g(x) + \int_{-\epsilon}^{\epsilon}dx \delta V g(x)) = 0, \label{ww}
\end{eqnarray}
where  $\delta V = [[V,h_0^{1/2}], h_0^{-1/2}]$.

We write $ \sqrt{p^2+m^2} = \sum_{n=0}^{\infty} a_n p^{2n}$, in terms of the coefficients $a_n$ obtained from the binomial expansion. The first term in Eq. (\ref{ww}) becomes
\begin{eqnarray}
&{}&\sum_{n=1}^{\infty} a_n \int_{-\epsilon}^{\epsilon}dx (-\partial_x^2)^n g(x) = \sum_{n=1}^{\infty} a_n (-1)^n (\partial_x^{2n-1}g(\epsilon) - \partial_x^{2n-1}g(-\epsilon) = \nonumber \\
&-&\sum_{n=1}^{\infty} a_n  \left( (k_2)^{2n-2}g'(\epsilon) - (k_1)^{2n-2}g'(-\epsilon) \right)=
-\frac{\sum_{n=1}^{\infty} a_n k_2^{2n}}{k_2^2} g'(\epsilon) + \frac{\sum_{n=1}^{\infty} a_n k_1^{2n}}{k_1^2} g'(-\epsilon) =
\nonumber \\
&-& \frac{\sqrt{m^2 +k_2^2} - m}{k_2^2}  g'(\epsilon) + \frac{\sqrt{m^2 +k_1^2} - m}{k_1^2}  g'(-\epsilon).
\end{eqnarray}
Writing $1/ \sqrt{p^2+m^2} = \sum_n b_n  p^{2n}$, the operator $\delta V$ can be expressed as a series
\begin{eqnarray}
\delta V = -V_0 \sum_n a_n \sum_m b_m \sum_{k=0}^{2n} \sum_{l=0}^{2m} \hat{p}^{k+l} \delta'(\hat{x}) \hat{p}^{2n+2m-l-k},
\end{eqnarray}
where $V_0$ is the potential jump at $x = 0$. Consider the integral of a single term in the series above
\begin{eqnarray}
\int_{-\epsilon}^{\epsilon}dx  \hat{p}^{k+l} \delta'(x) \hat{p}^{2n+2m-k-l} g(x) = (-i)^{2n+2m} \sum_{r=0}^{k+l}  \int_{-\epsilon}^{\epsilon}dx \left( \begin{array}{c} k+l \\ r \end{array} \right) \delta^{(1+r)}(x) \partial_x^{2n+2m-r}g(x) = \nonumber \\
=  (-i)^{2n+2m} \sum_{r=0}^{k+l} \left( \begin{array}{c} k+l \\ r \end{array} \right) (-1)^r \int_{-\epsilon}^{\epsilon}dx  \delta'(x) \partial_x^{2n+2m}g(x).
\end{eqnarray}
The sum over $r$ involves the term $\sum_{r=0}^{k+l} \left( \begin{array}{c} k+l \\r \end{array} \right) (-1)^r = (1-1)^r = 0$. Hence, the integral above vanishes, and so does the second term in Eq. (\ref{ww}).

Our final result is the junction condition
\begin{eqnarray}
\frac{\sqrt{m^2 +k_2^2} - m}{k_2^2}  g'(0_+) = \frac{\sqrt{m^2 +k_1^2} - m}{k_1^2}  g'(0_-), \label{jun}
\end{eqnarray}
which is to be implemented together with the continuity condition
\begin{eqnarray}
g(0_+) = g(0_-)
\end{eqnarray}
For $|k_1| <<m $ and $|k_2| << m$, Eq. (\ref{jun})  reduces to the non-relativistic junction condition $g'(0_+) = g'(0_-)$.

\subsection{Square barrier}
\subsubsection{Transmission amplitude}
Next, we consider the square barrier potential,
\begin{eqnarray}
V(x) = \left\{ \begin{array}{cc} V_0, & x \in [-\frac{d}{2}, \frac{d}{2}]\\
  0,  &  x \notin [-\frac{d}{2}, \frac{d}{2}] \end{array} \right.
\end{eqnarray}
For  $E - V_0 < m$, we apply the junction conditions (\ref{jun}) at $x = \pm \frac{d}{2}$, to obtain the transmission and reflection coefficients
\begin{eqnarray}
T_k = \frac{e^{-ikd}}{\cosh(\lambda_kd) - i \eta_k \sinh(\lambda_kd)} \label{tk1} \\
R_k = -i  \frac{e^{-ikd}\rho_k\sinh(\lambda_kd)}{\cosh(\lambda_kd) - i \eta_k \sinh(\lambda_kd)} \label{rk1}
\end{eqnarray}
where
\begin{eqnarray}
\lambda_k = \sqrt{m^2-(E-V_0)^2}, \label{lak}
\end{eqnarray}
 and the functions
\begin{eqnarray}
\eta_k = \frac{1}{2} \left(e_k - \frac{1}{e_k}\right)\\ \label{sik}
\rho_k = \frac{1}{2} \left(e_k + \frac{1}{e_k}\right) \label{rok}
\end{eqnarray}
are expressed in terms of the quantity

\begin{eqnarray}
e_k = \frac{\lambda_k}{k} \frac{\sqrt{m^2+k^2} - m}{m - \sqrt{m^2 - \lambda_k^2}}. \label{fk}
\end{eqnarray}
In the non-relativistic limit, $e_k = k/\lambda_k$ and Eqs. (\ref{tk1}---\ref{rk1}) reduce to the standard textbook expressions for the square barrier potential.

\subsubsection{Tunneling time}

Expressing the transmission amplitude (\ref{tk1}) as $T_k = |T_k|e^{i\phi_k}$, we note that the phase $\phi_k$ varies much faster with $k$ than $|T_k|$. Hence, when computing the probability density Eq. (\ref{parplt}) for a sufficiently narrow initial wave-packet, we may assume that $|T_k| \simeq |T_p|$.  The velocity $v_k$ varies slowly with $k$ in comparison to the phases, hence, we also set $v_k = v_p$. Expanding the phase factor to first order in $k - p$, $\phi_k = \phi_p + \phi'_k(k - p)$, we obtain
\begin{eqnarray}
P(L, t) = v_p |T_p|^2 |u_0[v_p t - (L + x_0 + \phi'_p)]|^2, \label{p1b}
\end{eqnarray}
where $u_0(x) = \int \frac{dk}{2 \pi} e^{-ikx} \tilde{u}_0(k)$ is the wave function $u_0$ in the position representation. The probability density Eq. (\ref{p1b}) exhibits a single peak at $ t = (L + x_0 + \phi'_p)$, thus leading to the identification of delay time  $t_d(p) = \phi'_p$, in accordance with Eq. (\ref{delt}). Using Eq. (\ref{tk1}), we obtain

\begin{eqnarray}
t_d(p) = - \frac{d}{v_p} + \tau_p,
\end{eqnarray}
where the tunneling time $\tau_k$ is
\begin{eqnarray}
\tau_p = \frac{-\eta_p \frac{\sqrt{m^2 - \lambda_p^2}}{\lambda_p} d + m \rho_p  (\frac{1}{p}+\frac{1}{\lambda_p}) \sinh (\lambda_p d) \cosh(\lambda_pd)}{\cosh^2(\lambda_pd) + \eta_p^2 \sinh^2(\lambda_pd)} \label{tauk}
\end{eqnarray}
Eq.  applies to all values of $p$ corresponding to tunneling: $0 \leq p \leq \sqrt{(m + V_0)^2 - m^2}$.

For a long barrier $e^{-\lambda_k d } << 1$,
\begin{eqnarray}
T_k = 2 \frac{e^{-ikd} e^{-\lambda_kd} }{1 - i \eta_k}, \label{tk}
\end{eqnarray}
and the tunneling time is independent of $d$
\begin{eqnarray}
\tau_p =   m \frac{\rho_p}{1 + \eta^2_p} (\frac{1}{p} + \frac{1}{\lambda_p}). \label{tauk2}
\end{eqnarray}
\begin{figure}[tbp]
\includegraphics[height=5cm]{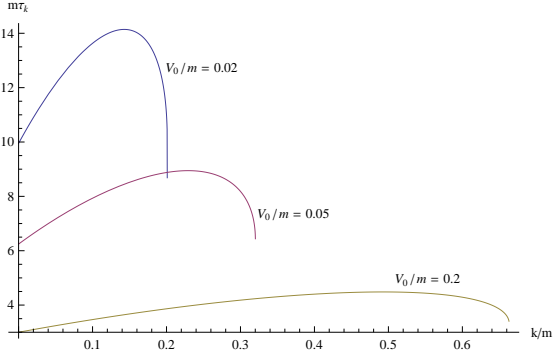} \caption{ \small Tunneling times $\tau_p$ as a function of the incoming momentum $p \in [0, \sqrt{2mV_0 + V_0^2}]$, for different values of the potential $V_0$ and barrier length $ d = 5000m^{-1}m$, $m$ the particle's mass. }
\end{figure}

\subsection{Double square barrier}

Next, we consider the case of the symmetric double square barrier, that is, the piecewise constant potential
\begin{eqnarray}
V(x) = \left\{ \begin{array}{cc} 0 & x \in (-\infty, -a - \frac{r}{2}) \cup (-\frac{r}{2}, \frac{r}{2}) \cup (a + \frac{r}{2}, \infty) \\
V_0 & x\in [-a - \frac{r}{2}, -\frac{r}{2}] \cup [\frac{r}{2}, a + \frac{r}{2}] \end{array} \right..
\end{eqnarray}
This potential corresponds to two identical square barriers of width $a$ separated by a distance $r$.

\subsubsection{The transmission amplitude}
We compute the transmission probability by implementing the junction conditions (\ref{jun}) at $x = \pm (a + \frac{r}{2})$ and at $x = \pm \frac{r}{2}$. We find,
\begin{eqnarray}
T_k = \frac{e^{-ik(r + 2a)}}{(\cosh(\lambda_k a) - i \eta_k \sinh(\lambda_k a))^2 e^{-ikr} + \rho_k^2 \sinh^2(\lambda_k a) e^{ikr}}, \label{tdouble}
\end{eqnarray}
where $\lambda_k$, $\eta_k $ and $\rho_k$ are functions of $k$ that refer to a single barrier and they are defined in Eqs. (\ref{lak}---\ref{rok}).

 The transmission amplitude Eq. (\ref{tdouble}) can also be expressed as
\begin{eqnarray}
T_k = \frac{T_{0k}^2}{1 - R_{0k}^2 e^{2ik(r+a)}} , \label{tdouble2}
\end{eqnarray}
where  $T_{0k}$ and $R_{0k}$ stand for the transmission and reflection amplitudes for the {\em single} barrier, as given by Eqs. (\ref{tk1}) and (\ref{rk1}) respectively.

Writing $T_{0k} = |T_{0k}|e^{i \phi_k}$ and $R_{0k} = - i |R_{0k}| e^{i \phi_k}$, Eq. (\ref{tdouble2}) becomes
\begin{eqnarray}
T_k = \frac{T_{0k}^2}{1 + |R_{0k}|^2 e^{2i[k(r+a)+\phi_k]}} \label{tdouble3}
\end{eqnarray}
Eq. (\ref{tdouble3}) implies that for $\cos\left[2(k(r+a) +\phi_k)\right] = -1$, $|T_k| = 1$. It follows that any  momenta $0 \leq k_n \leq \sqrt{V_0^2+ 2 mV_0}$ that are solutions to the equation \begin{eqnarray}
2 k_n (r+a) + \phi_{k_n} = (2n+1) \pi, \hspace{1cm} n = 0, 1, 2, \ldots \label{resonances}
  \end{eqnarray}
  correspond to {\em resonances} of the double-barrier potential.

\subsubsection{Large inter-barrier separation}

Next, we expand the denominator of the transmission amplitude Eq. (\ref{tdouble3}) as a geometric series,
\begin{eqnarray}
T_k = |T_{0k}|^2 \sum_{n=0}^{\infty} (-1)^n|R_{0k}|^{2n} e^{2i[nk(R+a)+(n+1)\phi_k]}. \label{tdouble4}
\end{eqnarray}

Substituting into Eq. (\ref{parplt}), we find that the   probability density $P(L, t)$ involves the amplitude
\begin{eqnarray}
{\cal A}(L, t) = \sum_{n = 0}^{\infty} (-1)^n \int_0^{\infty}dk  \sqrt{v_k} \tilde{u}_0(k-p) |T_{0k}|^2 |R_{0k}|^{2n} e^{ik(x_0+L)-iE_k t + 2i[n k (R + a) + (n + 1) \phi_k]}. \label{amplit1}
\end{eqnarray}
We evaluate the integral Eq. (\ref{amplit1}) using the same approximations as in Sec. 4.2.2. We set $|T_{0k}| = |T_{0p}|$, $|R_{0k}| = |R_{0p}|$, $v_k = v_p$, and we expand the phase factor to first order in $ k- p$.  Taking the limits of integration to $k \in (-\infty, \infty)$, we obtain
\begin{eqnarray}
{\cal A}(L, t) &=& \sqrt{v_p} |T_{0p}|^2 e^{ipL - iE_p t } \\ \nonumber
&\times& \sum_{n=0}^{\infty} (-1)^n |R_{0p}|^{2n} e^{2i[np(R+a)+(n+1)\phi_p]} u_0[(x_0 + L)-v_p t + 2 \phi'_p + 2n (r + v_p \tau_p)] . \label{ampl5b},
\end{eqnarray}
where  $\tau_p$ is the tunneling time, Eq. (\ref{tunt}), and $u_0(x)$ is the  wave function $\tilde{u}_0$ transformed to the position representation.

The amplitude Eq. (\ref{ampl5b}) involves a sum of terms labeled by $n$, with each term peaked around time
\begin{eqnarray}
t_n = [x_0 + L + \phi'_p + 2n (r + v_p \tau_p)]/v_p. \label{tn1}
\end{eqnarray}
 The distance between two successive peaks is
\begin{eqnarray}
\Delta t= 2(r/v_p + \tau_p). \label{deltat}
 \end{eqnarray}
 If the position spread $\sigma_x > 1/(2 \sigma_p)$ of the wave-function $u_0$ satisfies $\sigma_x << v_p \Delta t$, then there is no overlap between the terms in the sum Eq. (\ref{ampl5b}). Hence, the amplitude Eq. (\ref{ampl5b}) is close to zero except for neighborhoods of width $\sigma_x/v_p$ peaked around the values $t_n$, Eq. (\ref{tn1}). The first peak in the amplitude  is at time
\begin{eqnarray}
t_0 = (L + x_0)/v_p + 2 \phi'_p/v_p. \label{t0}
\end{eqnarray}
  The next  term in the series Eq. (\ref{ampl5b}) has a peak at $t_0 +\Delta t$. The delay by a factor $\Delta t$ can be attributed to the fraction of particles transmitted through the first barrier, reflected at the second barrier, reflected at the first barrier and exiting the second barrier in its second attempt. Similarly the peak at time $t_n$ corresponds to  particles that were transmitted through the first barrier, they were subsequently reflected $n$ times at the second barrier and $n$ times at the first barrier, and finally exited after their $(n+1)$-th attempt on the second barrier. The attempts of crossing the barrier are statistically independent: a particle exiting in its $(n+1)$-th attempt has been reflected $2n$ times. Since the amplitude for a single reflection is $R_{0p}$, the contribution of the successful exit by the $(n+1)$-th attempt is   suppressed by a factor $|R_{0p}|^{2n}$.

The detection probability $P(L,t) = |{\cal A}(L, t)|^2$ is
\begin{eqnarray}
P(L,t) = v_p |T_{0p}|^4 \sum_{n=0}^{\infty} \sum_{m=0}^{\infty} |R_{0p}|^{2(n+m)} e^{i (n-m)\beta_p} u_0(v_p \Delta t [n - (t-t_0)/\Delta t])u_0(v_p \Delta t [m - (t-t_0)/\Delta t]), \label{pltdsum}
\end{eqnarray}
where

\begin{eqnarray}
\beta_p = (2 p(r + a) + 2 \phi_p + \pi) \mod(2 \pi) \label{lap}
\end{eqnarray}

For $\sigma_x << v_p \Delta t$, the function $u_0$ in Eq. (\ref{pltdsum})  approximates well a delta function. Delta functions satisfy the property $\delta(x) \delta (y) = \delta\left(\frac{x+y}{2}\right) \delta(x-y)$. We expect that an analogous equality also holds approximately for smeared delta functions, i.e., for families of functions $F_a(x)$, such that  $\lim_{a \rightarrow 0} F_a(x) = \delta (x)$, when $a$ is close to zero. Hence, we can write $u_0(x) u_0(y) = u_0[\frac{1}{2}(x+y)] u_0(x-y)$ in Eq. (\ref{pltdsum}); the equality is {\em exact} for Gaussian wave-functions.
   We also rearrange the double sum in Eq. (\ref{pltdsum}) as $\sum_{N = 0}^\infty \sum_{M = -N}^N$, where $N = m+n$ and $N = n -m$. Then, the  time dependence of Eq. (\ref{pltdsum}) is contained in a term

\begin{eqnarray}
 \sum_{N=0}^{\infty} |R_{0p}|^{2N} u_0(v_p \Delta t[\frac{1}{2}N - (t-t_0)/\Delta t]). \label{peaks}
 \end{eqnarray}

Thus, the probability distribution $P(L,t)$ is an infinite sum of terms, each sharply  peaked at times $t_N = t_0 + \frac{1}{2}N \Delta t$ and suppressed by a factor $|R_{0p}|^{2N}$. If the time-scale of observation is much larger than $\Delta t$ and $|T_{0p}|^2 << 1$, we can approximate the probability distribution by a smooth function connecting the peaks of Eq. (\ref{peaks}), modulo an overall normalization constant. In this approximation, the probability distribution $P(L,t)$ vanishes for $t < t_0$. For $t > t_0$, $P(L, t)$ is proportional to $|R_{p0}|^{2N} = e^{N \log(1 - |T_{0p}|^2)} \simeq e^{-N |T_{0p}|^2}$, where $N = 2 (t -t_0)/\Delta t$.

Hence,

\begin{eqnarray}
P(L,t) = \left\{ \begin{array}{cc} 0& t < t_0 \\
 C e^{-\Gamma_p(t-t_0)} & t > t_0 \end{array} \right., \label{plt6}
\end{eqnarray}
where
\begin{eqnarray}
\Gamma_p =  \frac{2 |T_{0p}|^2}{\Delta t} = \frac{|T_{0p}|^2}{\frac{r}{v_p} + \tau_p}. \label{gap}
\end{eqnarray}
is a decay coefficient associated with the barrier, and $C$ is a positive constant.

Eq. (\ref{plt6}) applies for (i) $p$ far from a resonance frequency of the  double barrier system and (ii) for wave-packets with  position spread much smaller than $v_p\Delta t$.

\begin{figure}[tbp]
\includegraphics[height=5cm]{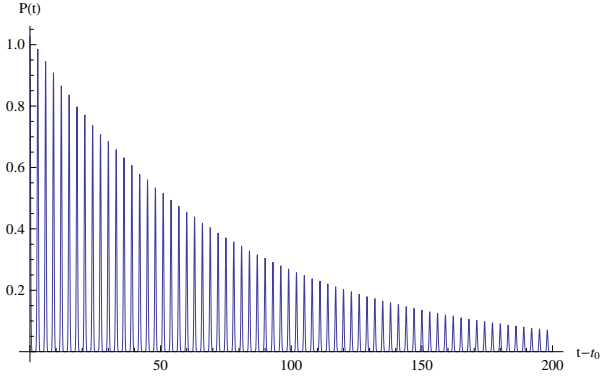} \caption{ \small The  probability density $P(L,t)$, Eq. (\ref{pltdsum}), for particle detection for the double square barrier potential, as a function of the time $t - t_0$ after first detection. The probability density is a sum of peaks separated by distance $\Delta t$. The peaks in the distribution decay exponentially with rate $\Gamma_p$, Eq. (\ref{gap}). }
\end{figure}

The coefficient $C$ Eq. (\ref{plt6}) is determined by the requirement that the time-integrated probability is consistent with Eq. (\ref{tint}), i.e., the requirement that

\begin{eqnarray}
\int_0^{\infty} P(L,t) dt \simeq |T_p|^2 =\frac{|T_{0p}|^4}{1 + |R_{0p}|^4 + 2 |R_{0p}|^2 \cos \beta_p}.
\end{eqnarray}
This leads to
 \begin{eqnarray}
P(L,t) = \left\{ \begin{array}{cc} 0& t < t_0 \\
  |T_p|^2 \Gamma_p e^{-\Gamma_p(t-t_0)} & t > t_0 \end{array} \right., \label{plt7}
\end{eqnarray}

The physical interpretation of Eq. (\ref{plt7}), expressed in classical language, is the following.
 The first signal arrives at the detector at a time $t_0$ which corresponds to a delay time $ t_d(p) = 2 \phi'_p /v_p$ that is twice the delay time of the single barrier. The associated tunneling time $\tau^{2b}_p$ for the particles at first detection is
\begin{eqnarray}
\tau^{2b}_p = t_d(p) + \frac{r +2a}{v_p} = 2 \tau_p + \frac{r}{v_p}. \label{t2b}
\end{eqnarray}
  The classical interpretation of Eq. (\ref{t2b}) is that the particles cross the two barriers independently and at time equal to the tunneling time $\tau_p$ of each barrier. In the inter-barrier region, the particles propagate freely with velocity $v_p$. Particles crossing the first barrier, are trapped in the inter-barrier region, and escape only after multiple attempts to cross the second barrier. The detection probability of those later-escaping particles decays exponentially with a rate $\Gamma_p$.

\subsubsection{Wave-packets narrowly concentrated in momentum}
In Sec. 4.3.3,  we considered the regime $\sigma_p v_p \Delta t >> 1$ that corresponds to an inter-barrier distance $r$ much larger than the spread of the initial wave-packet. In the opposite regime, of small inter-barrier distance $r$, the peaks in the amplitude Eq. (\ref{ampl5b}) essentially overlap. Thus, it is a meaningful approximation to substitute the infinite sum in Eq. (\ref{ampl5b}) with an integral over a continuous variable, using the Euler-MacLaurin summation formula.

We consider a Gaussian  $\tilde{u}_0(k)$,
\begin{eqnarray}
\tilde{u}_0(k) = \left(\frac{4 \pi}{\sigma_p^2}\right)^{1/4} e^{-\frac{k^2}{4\sigma_p^2}}. \label{Gauss}
\end{eqnarray}
Then, Eq. (\ref{ampl5b}) becomes
\begin{eqnarray}
{\cal A}(L, t) \simeq \sqrt{v_p} |T_{0p}|^2 \left(\frac{\sigma_p^2}{\pi}\right)^{1/4}e^{i(pL-E_pt+2\phi_p)}  \left[ \int_0^{\infty} dn e^{ - \sigma_p^2 v_p^2 \Delta t^2 \left(n - \frac{t - t_0}{\Delta t}\right)^2 - n \left(|T_{0p}|^2 - i \beta_p\right)} + \frac{1}{2} e^{-\sigma_p^2 v_p^2(t-t_0)^2}\right], \label{ampl7}
\end{eqnarray}
where $t_0$ is given by Eq. (\ref{t0}), $\Delta t$ is given by Eq. (\ref{deltat}) and $\beta_p$ by Eq. (\ref{lap}).

For $t < t_0$, the amplitude Eq. (\ref{ampl7}) is strongly suppressed. In the vicinity of $t = t_0$, small values of $n$ dominate in the integral Eq. (\ref{ampl7}), so we can ignore the contribution of the term $e^{-n |T_{0p}|^2}$. Then,
\begin{eqnarray}
P(L,t) = v_p |T_{0p}|^4 \left(\frac{\sigma_p^2}{\pi}\right)^{1/2} \left|\frac{\sqrt{\pi}}{2 \sigma_p v_p \Delta t} e^{-\mu_p^2} \mbox{erfc}(-\sigma_p v_p (t-t_0)+ i \mu_p) + \frac{1}{2} e^{-\sigma_p^2 v_p^2(t-t_0)^2} \right|^2, \label{plt11}
\end{eqnarray}
where $\mu_p = \frac{\beta_p}{2 \sigma_p v_p \Delta t}$, and $\mbox{erfc}$ stands for the complementary error function
\begin{eqnarray}
\mbox{erfc}(x) = \frac{2}{\sqrt{\pi}}\int_x^{\infty} e^{-t^2}dt.
\end{eqnarray}

The probability density (\ref{plt11}) is strongly peaked around $t = t_0$, hence, also in this regime, the delay time   for the double barriers is twice the delay time of the single barrier.

At later times ($ t - t_0 >> 1/(\sigma_p v_p)$),  we can take the limit of integration in Eq. (\ref{ampl7}) to $- \infty$, and drop the term outside the integral. Then,
\begin{eqnarray}
P(L,t) = K e^{-\Gamma_p(t-t_0)},
\end{eqnarray}
where
\begin{eqnarray}
K = \frac{\sqrt{\pi} \Gamma_p^2}{4 v_p \sigma_p} e^{\frac{\Gamma_p^2}{16} - \mu_p^2}
\end{eqnarray}
is a constant.

Hence, following a transient period of order $1/(\sigma_p v_p)$ after the first particle detection, the probability $P(L,t)$  decays exponentially with rate $\Gamma_p$. This qualitative behavior is verified by a numerical evaluation of the probability density Eq. (\ref{pltdsum}) in the regime $\sigma_p v_p \Delta t \sim 1$---see, Fig. 3.
\begin{figure}[tbp]
\includegraphics[height=5cm]{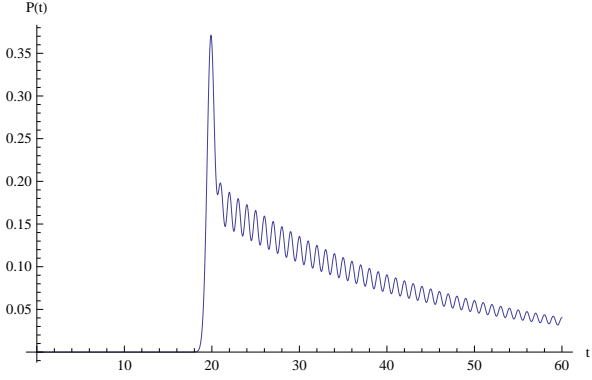} \caption{ \small The  probability density $P(L,t)$, Eq. (\ref{pltdsum}), as a function of the time $t$, in the regime $\sigma_p v_p \Delta t \sim 1$. The probability exhibits a sharp peak at the time $t_0$ of first detection, followed by a sequence of shorter peaks that decay exponentially with rate $\Gamma_p$, given by Eq. (\ref{gap}). }
\end{figure}

\subsubsection{Resonant tunneling}
The results of Secs. 4.3.2 and 4.3.3 rely on the assumption that the particles' momentum $p$ is far from the resonances of the double barrier, Eq. (\ref{resonances}). Here, we consider the probability distribution $P(L,t)$ for an initial state centered on a momentum $p$ that is close to a resonance frequency $k_0$.

Near the resonance $k = k_0$, the transmission amplitude is approximated by
\begin{eqnarray}
T_k = \frac{|T_{0k_0}|^2 e^{2 i \phi_{k}}}{1 -|R_{0k_0}|^2 e^{2 i(k-k_0) (r+a + \phi'_{k_0})}}.
\end{eqnarray}
For $|T_{0k_0}|<<1$, the  modulus square $|T_k|^2$ is of the order of $|T_{0k_0}|^4$ except for a narrow strip of momenta around $k_0$ where $|T_k|^2$ approaches unity. These momenta will dominate in the probability distribution $P(L, t)$. Hence, we expand the phase factor $e^{2 i(k-k_0) (r+a + \phi_{k_0}')}$, keeping only  first order terms with respect to $k - k_0$. The transition amplitude becomes
\begin{eqnarray}
T_k = \frac{e^{2 i \phi_k } }{1 - i \frac{2v_{k_0}}{\Gamma_{k_0}} (k-k_0)},
\end{eqnarray}
 where $\Gamma_{k_0}$ is the decay coefficient, Eq. (\ref{gap}), evaluated at the resonance frequency.

 In this approximation, the squared amplitude
 \begin{eqnarray}
 |T_k|^2 = \frac{1}{1 +  \frac{2v_{k_0}^2}{\Gamma_{k_0}^2} (k-k_0)^2}. \label{tklor}
 \end{eqnarray}
is a Lorentzian  of width $\Gamma_{k_0}/(2v_{k_0})$ centered around $k = k_0 $.

We consider a initial wave-packet centered around momentum $k = p$, such that $p$ lies within the peak of the Lorentzian Eq. (\ref{tklor}). The probability density $P(L,t)$ is obtained from the squared modulus of the amplitude
\begin{eqnarray}
{\cal A}(L,t) = \int_0^{\infty} \frac{dk}{2\pi} \sqrt{v_k}\frac{1}{1 - i \frac{2v_{k_0}}{\Gamma_{k_0}} (k-k_0)}\tilde{u}_0(k-p)e^{i(k(x_0+L) - E_k t + 2 \phi_k)}
\end{eqnarray}
We assume that the momentum spread $\sigma_p$ of $\tilde{\phi}_0$ is much smaller than $p$, so that we can enlarge the integration range to $(-\infty, \infty)$. We also expand the phase factor to first order in $q = k - p$, to obtain
\begin{eqnarray}
{\cal A}(L,t) = \sqrt{v_p} e^{ip(x_0+L) -iE_pt + i \phi_p} \int_{-\infty}^{\infty} \frac{dq}{2\pi} \frac{1}{1 - i \frac{2v_{k_0}}{\Gamma_{k_0}} [q- (k_0-p))} \tilde{u}_0(k-p) e^{-iyv_p(t -t_0)}. \label{ampl9}
\end{eqnarray}

We  calculate the integral in Eq. (\ref{ampl9}) analytically, by choosing a Lorentzian wave-packet
\begin{eqnarray}
\tilde{u}_0(k) = \frac{1}{\sqrt{\sigma_p}} \frac{1}{1 + \frac{k^2}{\sigma_p^2}}. \label{lorentz}
\end{eqnarray}
Then, we obtain
\begin{eqnarray}
{\cal A}(L,t) &=&   \frac{\frac{i \Gamma_{k_0}\sqrt{v_p}}{2 \sqrt{\sigma_p} v_{k_0} } e^{ip(x_0+L) -iE_pt + i \phi_p} }{1 + \left(\frac{k_0 - p}{\sigma_p} - i \frac{\Gamma_{k_0}}{2v_{k_0} \sigma_p}\right)^2} \nonumber \\
&\times&
  \left\{ \begin{array}{cc} \left[\frac{k_0-p}{\sigma_p} -i \left(\frac{\Gamma_{k_0}}{2v_{k_0} \sigma_p}  - 1 \right)\right] e^{-v_p \sigma_p |t -t_0|},&  t < t_0 \\
-2i e^{-i v_p(k_0 - p) (t-t_0) -\frac{\Gamma_{k_0}}{2}|t-t_0|} + \left[\frac{k_0-p}{\sigma_p} -i\left( \frac{\Gamma_{k_0}}{2v_{k_0} \sigma_p}  +1 \right) \right] e^{-v_p \sigma_p |t -t_0|}, & t > t_0\end{array} \right. \label{ampl10}
\end{eqnarray}
We note that the amplitude is suppressed for times $t < t_0$.   The first detection signal appears around $t = t_0$. Again, this implies that the time delay due to the barrier is $2 \phi'_p/v_p$.

The probability density $P(L,t) = |{\cal A}(L, t)|^2$ is computed straightforwardly. In the regime of exact resonance ($|k_0 - p| << \sigma_p$), and for $\Gamma_{k_0} >> \sigma_p v_p$, it simplifies

\begin{eqnarray}
P(L,t) \simeq \frac{\Gamma_{k_0}^2 v_p}{\sigma_p v_{k_0}^2}    \left\{ \begin{array}{cc} e^{-v_p \sigma_p |t -t_0|} & t < t_0 \\
4 e^{-\Gamma_{k_0}(t - t_0) } + e^{- 2 \sigma_p v_p(t -t_0)} + 4 e^{- (\frac{\Gamma_{k_0}}{2} + \sigma_p v_p)(t-t_0) } \cos[v_p(k_0 - p) (t-t_0)], & t > t_0. \end{array} \right.
\end{eqnarray}
The instant $t_0$ of first detection is followed by a transient regime, and at times $t$ such that $t - t_0 >> (\sigma_p v_p)^{-1}$, the probability decays purely exponentially with  rate $\Gamma_{k_0}$.

Hence, also in the case of resonance the double square barrier is characterized by two time-scales: the delay time $2 \phi'_p/v_p$ and the  decay rate $\Gamma_{k_0}$ of the detection probability at later times.
\begin{figure}[tbp]
\includegraphics[height=5cm]{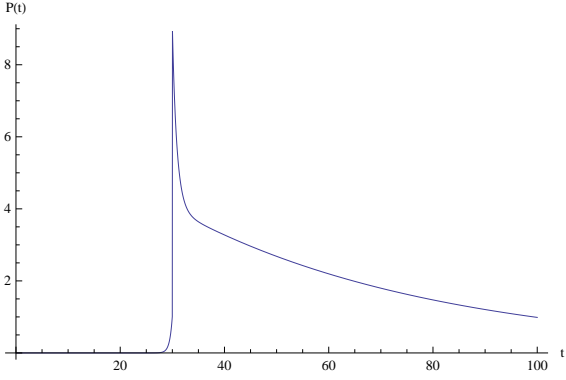} \caption{ \small The  probability density $P(L,t)$, Eq. (\ref{pltdsum}), as a function of the time $t$, near resonance and for an initial Lorentzian wave-packet, Eq. (\ref{lorentz}). The probability exhibits a sharp peak at the time $t_0$ of first detection, and an exponential decay phase with rate $\Gamma_{k_0}$, given by Eq. (\ref{gap}). }
\end{figure}
\subsubsection{Multiple resonances}
For large inter-barrier distances, the difference between two successive resonance momenta is of the order of  $\pi/r$. Hence, for wave packets with significant momentum spread $\sigma_p \sim \pi/r$, two or more resonances may contribute significantly to the detection probability. In this case, we express the transmission amplitude as
\begin{eqnarray}
T_k = e^{2 i \phi_k } \sum_n \frac{1 }{1 - i \frac{2v_{k_n}}{\Gamma_{k_n}} (k-k_n)},
\end{eqnarray}
where $k_n$ correspond to the resonance frequencies.

Using the approximations employed in the derivation of Eq. (\ref{ampl9}, we obtain
\begin{eqnarray}
{\cal A}(L,t) = \sqrt{v_p} e^{ip(x_0+L) -iE_pt + i \phi_p} \sum_n \int_{-\infty}^{\infty} \frac{dq}{2\pi} \frac{1}{1 - i \frac{2v_{k_n}}{\Gamma_{k_n}} [q- (k_n-p))} \tilde{u}_0(k-p) e^{-iqv_p(t -t_0)}. \label{ampl11}
\end{eqnarray}
The amplitude ${\cal A}(L,t)$ is a sum of partial amplitudes ${\cal A}_n(t, L)$, each incorporating the contribution from a single resonance. For  the Lorentzian initial state Eq. (\ref{lorentz}), each amplitude ${\cal A}_n(L,t)$ is of the form (\ref{ampl10}). The resulting probability distribution has the same structure as Eq. (\ref{ampl10}) around $t = t_0$, since $t_0$ depends only on the mean momentum of the wave-packet. However, at later times the behavior differs. To a first approximation, the decay constants at different channels are equal $\Gamma_{k_n} \simeq \Gamma_p$. Then, the dominant contribution to $P(L,t)$ at later times is
\begin{eqnarray}
P(L,t) \sim |\sum_n e^{-iv_p(k_n-p)(t-t')}|^2 e^{-\Gamma_{p}(t-t_0)}.
\end{eqnarray}
 Thus, at later times, the exponential decay of the probability density $P(L,t)$ is modulated by oscillations of frequencies $v_p(k_n - k_m) \sim v_p/r$.

\begin{figure}[tbp]
\includegraphics[height=5cm]{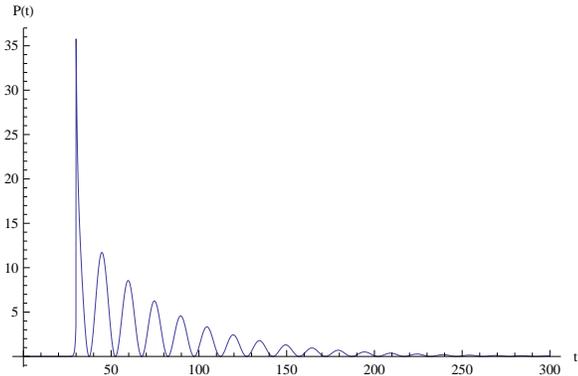} \caption{ \small The  probability density $P(L,t)$, Eq. (\ref{pltdsum}), as a function of the time $t$,  for an initial Lorentzian wave-packet, Eq. (\ref{lorentz}), that overlaps with two momentum resonances $k_1$ and $k_2$. The probability exhibits a sharp peak at the time $t_0$ of first detection, and  an exponential decay phase with rate $\Gamma_{k_1} \simeq \Gamma_{k_2}$, modulated by oscillations of frequency $\simeq v_p |k_2 - k_1|$. }
\end{figure}

\subsubsection{Summary}
We calculated the probability density $P(L,t)$ for the double-square barrier in different regimes and using different approximation schemes. In all regimes, $P(L,t)$ has the same qualitative behavior: it is characterized by a sharp peak at the time $t_0$ of first detection and then (sometimes after a transient period) decays exponentially to zero with a rate $\Gamma_p$. The probability distribution is determined by two parameters that are defined solely in terms of the double barrier's characteristics and of the incoming particle momentum: the delay time $t_d(p)$ associated with first detection, and the decay rate $\Gamma_p$. The delay time $t_d(p)$ is twice the delay time for a single barrier. From the value of the delay time, one is led to a classical picture of tunneling, according to which  particles cross the two barriers independently, each crossing taking time $\tau_p$, while the particles propagate freely in the inter-barrier region.

The exponential decay in the probability density $P(L,t)$ at late times is due to the fraction of particles that cross the first barrier, and then they are trapped in the inter-barrier region. For these particles, the double barrier acts as a potential well, which they can only exit through tunneling. The exponential decay law at later times is of the same origin as the exponential decay of unstable states through tunneling. Indeed, its derivation in Sec. 4.3.2 is rather similar to the classic studies of $\alpha$-decay by Gamow \cite{Gamow} and Gurney and Condon \cite{GuCo}---see also, Ref. \cite{An08}.

A key point that emerges from this analysis, is that the temporal aspects of tunneling cannot be captured by a single variable. The full construction of the probability distribution $P(L, t)$ is necessary. In particular, if one attempts to describe tunneling in terms of the time-delay obtained through a stationary phase approximation, as in Eq. (\ref{delt}), one is led to very different conclusions, namely, that  there exists a regime, in which the inferred tunneling time is independent of the inter-barrier distance $r$, a property referred to as the "generalized Hartmann effect" \cite{genhar, genhar2}. However, as shown in Ref. \cite{KuMa}, a simple stationary phase approximation does not capture the properties of the Schr\"odinger-evolved wave function in the double barrier system.  A careful implementation of the stationary phase approximation in the amplitude ${\cal A}(L,t)$ of Eq. (\ref{ampl5b}) leads to the results presented here.

\section{Eliminating the superluminality paradox}

In this section, we propose a resolution of the superluminality paradox in tunneling. Our proposal is based on the definition of the probability distribution $P(L,t)$ in terms of the correlation functions of a local QFT.

Eq. (\ref{delt}) for the time delay $t_d(k)$ applies only to regimes at which the saddle point approximation to $P(L,t)$ is meaningful. This includes several important cases, such as tunneling through a square barrier. As we showed in Sec. 4.2, the delay time for a square barrier coincides with the standard phase delay time of Bohm and Wigner \cite{BW}. In particular,  $t_d(k)$ saturates for sufficiently long barriers: $t_d(k) = -d/v_k + F(k)$.
  If one defines the time $\tau$ that the particle spent inside the barrier as $\tau = t_d + d/v_k$, then $\tau$ is independent of the barrier length $d$. This implies superluminal traversal velocity for   long barriers.

The problem with this line of reasoning is that the time delay  $t_d(k)$   is not a genuine quantum observable. It does not correspond to a self-adjoint operator (or a positive-operator-valued measure), and it  does not correspond a measurement record in the experiment. The measurement records correspond to the time of arrival $t$, not to $t_d$. The time delay is a parameter of the probability distribution for the detection times, and it can only be identified after the probability $P(L,t)$ has been determined.

In fact, the identification of the time delay  $t_d(k)$ requires the combination of data from two different time-of-arrival experiments, one with and one without the barrier. Each experiment records a probability distribution for the arrival times and $t_d(k)$ is identified as the difference of the corresponding mean values. For this reason, an inference of superluminality from the values of the delay time $t_d(k)$  involves an inadmissible treatment of measurement records that violates the complementarity principle. In no experiment is an actual superluminal signal recorded. The events of particle emission and particle detection are time-like separated. Their proper distance $\Delta s^2 = [(L+x_0)/v_k + t_d]^2 - (L + x_0)^2$ is positive in the regime $L >> d$ where Eq. (\ref{delt}) applies. This is because for $L >> d$ the particle spends considerably more time outside the barrier region.

The positivity of  $\Delta s^2$ for $L >> d$ relies on the fact that for the square barrier,
\begin{eqnarray}
t_d(k) > - d/v_k.  \label{bound}
 \end{eqnarray}
 However, the conclusion is not restricted to this particular potential. For $L >> d$, tunneling can be described as a scattering process, and the derivatives of the scattering phase shifts must have a lower bound in order for the $S$-matrix to be compatible with causality \cite{phshift}. The lower bound in the phase shifts implies a lower bound to the delay time $t_d(k)$. For non-relativistic particles and in absence of bound states,  the lower bound Eq. (\ref{bound}) applies \cite{mugashift}. In general, the corrections to this lower bound are of the order of the particle's de-Broglie wave-length divided by $v_p$, i.e., they are microscopic. They do not affect the argument about the time-like separation between the events of particle emission and particle detection.

Thus, the only conceivable way of recording a superluminal signal would be in an experimental set-up where both emitter and detector are placed very close to the barrier so that $\Delta s^2 <0$. This implies that $ r/v_p + t_d(p) < r$, where we write $r = L+ x_0$ for the source-detector distance. Since $t_d(p) > -d/ v_p$, the necessary condition for $\Delta s^2 < 0$ is that $d < r < r/(1 - v_p)$. The velocity $v_p$ is bounded above by $\sqrt{1 - (1 + \frac{V_0}{m})^{-2}}$: for larger velocities there is no tunneling.  The approximation of a background electric field fails for $V_0 > m$, because in this case the Hamiltonian Eq. (\ref{hamilton}) would possess negative-energy eigenstates. This implies an upper bound to the velocity  $v_p < \sqrt{3}/2 \simeq 0.87$. A less stringent upper bound to the velocity ($v_p < 2\sqrt{2}/3 \simeq 0.94$) follows from the requirement that the particle's kinetic energy is less than $2m$, so that particle-antiparticle pairs are not created spontaneously. 

Thus, we conclude that the only conceivable range of values for $r$ where a superluminal might be possible on the basis of Eq. (\ref{delt}) is
 \begin{eqnarray}
 d < r < 7.5 d,
 \end{eqnarray}
 i.e., the   emitter-detector distance $r$ is of the order of the barrier length $d$. However, in this case the  stationary phase approximation that leads to the delay time Eq. (\ref{delt}) is invalid.    To see this, note that the spread $\sigma_x$ in position of the initial wave-packet must be much smaller than $r$. For $r \sim d$, this implies that $\sigma_x >> d$. Hence, the momentum spread $\sigma_p$ must be very large, $\sigma_p >> 1/d$. This condition that is incompatible with the stationary phase approximation employed in the derivation of Eq. (\ref{delt}). The probability distribution $P(L, t)$ is strongly deformed and there is no meaningful definition of a delay time $t_d(k)$.

In fact, the absence of superluminal signals is guaranteed by Eq. (\ref{prob1b}) that relates the probability distribution $P(L, t)$, to the correlation functions of a local quantum field. A single-particle  state $| \psi_0 \rangle $ can be expressed as $| \psi_0 \rangle = \int dx \hat{\phi}_1(x) f^*(x)|0\rangle$, for some localized function $f(x)$. For a single-particle initial state centered around $x = -x_0 < -d/2$, Eq. (\ref{prob1b}) implies that the probability $P(L, t)$ is proportional to the four-point function
 \begin{eqnarray}
 \langle 0 | \hat{\phi}_1(-x_0,0) \hat{\phi}_1^{\dagger}(L,t) \hat{\phi}_1(L,t) \hat{\phi}_1^{\dagger}(-x_0,0)|0\rangle, \label{4pt}
 \end{eqnarray}
 modulo spatial smearing at the points of emission and detection. In Eq. (\ref{4pt}), $\hat{\phi}_1$ stands for the positive-frequency part of the Heisenberg field operator $\hat{\phi}$.

 Since the points $x = -x_0$ and $x = L$ are outside the barrier region, the field $\hat{\phi_1}$ is effectively free at these points, and they relate to the asymptotic fields defined at $t \rightarrow \pm \infty$ via evolution through the free-field Hamiltonian. In particular, $\hat{\phi}_1(-x_0, 0)$ corresponds to the in-field and $\hat{\phi}_1(L, t)$ to the out field of the $S$-matrix formalism. For a Hamiltonian quadratic to the quantum fields, the four point function Eq. (\ref{4pt}) factorizes into a product of two point functions. Hence,
 \begin{eqnarray}
 P(L, t) \sim | \Delta(-x_0, L; t)|^2 \label{estim}
 \end{eqnarray}
 where $\Delta(x,x';t) = [\hat{\phi}_1(L,t), \hat{\phi}_1^{\dagger}(-x_0, 0)] = \langle 0|\hat{\phi}_1(L,t) \hat{\phi}_1^{\dagger}(-x_0, 0)|0\rangle$. 
 
 Eq. (\ref{estim}) implies that the probability $P(L,t)$ involves propagation of the initial state through the two-point function $\Delta(x,x';t)$ of the quantum field. In any causal QFT, this Green's function, constructed from the $in$ and $out$ fields, {\em must vanish outside the light-cone}.  Otherwise, it would lead to a non-causal S-matrix.  It follows that the detection probabilities cannot involve superluminal signals. Thus, we claim that the QFT description of tunneling time, enabled through the use of the QTP method, eliminates the superluminality paradox.

The same point of principle holds for the electromagnetic analogues of tunneling, mentioned earlier, where superluminal group velocities have been recorded.  The analogy of those experiments to quantum tunneling is based on the following correspondence. The classical  Helmholtz equation for an electromagnetic field mode $\tilde{E}_{\omega}$ at frequency $\omega$ in an inhomogeneous dielectric medium of refraction index $n(x)$ is
\begin{eqnarray}
\partial_x^2 \tilde{E}_{\omega} + \left(n(x)\omega\right)^2 \tilde{E}_{\omega} = 0.
\end{eqnarray}
This is formally analogous to the time-independent Schr\"odinger equation in presence of a potential
\begin{eqnarray}
\partial_x^2 \psi + \left[2m(E-V(x))\right] = 0,
\end{eqnarray}
if we set $n(x) \omega = \left[2m(E-V(x))\right]^{1/2}$. Depending on the dielectric, it
is possible to have evanescent waves, which decay very much like the quantum mechanical wave functions in tunneling.

However, the analogy between the Helmholz and the Schr\"odinger equations holds only if they are viewed as describing classical waves equations. At the quantum level there is a significant difference. Time evolution according to Schr\"odinger's equation is always unitary. The effective Hamiltonian operator for the quantum electromagnetic field in presence of a   dielectric is
\begin{eqnarray}
\hat{H}_{eff} = \sum_a n(\omega_a) \omega_a \hat{a}^{\dagger}_a \hat{a}_a, \label{hem}
\end{eqnarray}
where $\hat{a}, \hat{a}^{\dagger}$ are creation and annihilation operators of the electromagnetic field, and $a$ labels the field modes. Evanescent waves appear for imaginary values of the refraction index $n(\omega_a)$, in which case the operator (\ref{hem}) becomes non-hermitian. The effective quantum evolution according to Eq. (\ref{hem}) would then be non-unitary. This is to be expected, since the microscopic mechanism generating an imaginary refraction index is the absorption of photons by the atoms of the medium. This implies that the electromagnetic analogues of tunneling are in fact analogues of time-of-arrival experiments in {\em open} quantum systems, where the effects of dissipation and noise have to be incorporated in the quantum description.

A full treatment of this effect requires the application of the QTP method to temporal measurements in open quantum systems, which will be the topic of a different publication.
Here, we point out that a fully quantum treatment of the electromagnetic field in inhomogeneous media requires the complete Quantum Electrodynamics (QED) Hamiltonian for the interaction between the electromagnetic field and the medium. That is, we consider a Hilbert space ${\cal H}_{tot} = {\cal H}_{EM} \otimes {\cal H}_{med} \otimes {\cal H}_{app}$, where ${\cal H}_{EM}$ is associated with the electromagnetic field, ${\cal H}_{med}$ to the particles forming the dielectric medium and ${\cal H}_{app}$ associated with the measurement apparatus. 

The dynamics on ${\cal H}_{EM} \otimes {\cal H}_{med}$ is governed by the QED Hamiltonian
\begin{eqnarray}
\hat{H}_{QED} = \hat{H}_{EM} \otimes 1 + 1 \otimes \hat{H}_{med} + \int dx \hat{A}^i(x)\otimes \hat{j}_i(x),
\end{eqnarray}
where $\hat{H}_{EM} $ is the Hamiltonian for the free electromagnetic field, $\hat{H}_{med}$ is the self-Hamiltonian for the particles in the dielectric medium, $\hat{A}^i(x)$ the electromagnetic potential and $\hat{j}_i$ the current associated with the medium.

The coupling of the electromagnetic field with the detector is governed by an interaction Hamiltonian
\begin{eqnarray}
\hat{H}_I = \int dx \hat{A}^i(x) \hat{J}_i(x),
\end{eqnarray}
where $\hat{J}_i(x)$ is the electric current associated with the detector degrees of freedom. We can then derive an equation for the probability density of detection $P(L,t)$ analogous to Eq. (\ref{prob1b})
\begin{eqnarray}
P(L, t) = \int d\tau \int dz R^{ij}(z, \tau)   \langle \psi_0, \Omega|\hat{A}_i^{\dagger}(L -\frac{z}{2}, t-\frac{\tau}{2}) \hat{A}_j(L+\frac{z}{2}, t+\frac{\tau}{2})|\psi_0, \Omega\rangle, \label{prob1f}
\end{eqnarray}
where $R^{ij}(\tau, z)$ is a kernel incorporating the detector degrees of freedom, and $|\psi_0, \Omega \rangle = |\psi_0\rangle \otimes |\Omega\rangle$, where $|\psi_0 \rangle$ is the initially prepared electromagnetic field state, and $|\Omega \rangle$ the initial state of the particles in the medium (for example, an energy eigenstate).
Using same arguments as the ones leading to  Eq. (\ref{estim}) for $P(L,t)$, we can show that the propagation of the electromagnetic field signal in Eq. (\ref{prob1f}) is guided by the two-point function of the field $\hat{A}_i$ that corresponds to photon emission and detection. If QED is a consistent quantum field theory, these functions must be causal, irrespective of   the initial state $|\Omega\rangle $ of the particles in the dielectric medium.
 Thus, no superluminal signal is to be expected in time-of-arrival measurements of photons through an absorbing medium.

To summarize, our proposed resolution to the superluminality paradox is the following. The time delay $t_d(k)$ can indeed be determined in {\em some} time-of-arrival experiments on tunneling set-ups. However, it is not a genuine quantum observable, but a parameter of the probability distribution $P(L,t)$. An inference of superluminal velocities in the tunneling region (or indeed any velocity) from the value of $t_d(k)$ involves classical reasoning that is incompatible with the rules of quantum theory. All physical information about signal propagation is contained in the  probability distribution $P(L,t)$, Eq. (\ref{prob1}). This is guaranteed to be causal, because, the QTP method enables its definition  in terms of the correlation functions of a local QFT.

In this sense, the apparent superluminality paradox in tunneling is a non-trivial manifestation of the particle-wave duality in quantum theory. Group velocities may be greater than the speed of light, but they are defined in terms of set-ups that measure wave properties of the quantum field, and not arrival  times. The latter correspond to the particle aspects of the quantum field, and they are measured in different experiments. Inferences of superluminality follow from attempts to relate observables defined in different experiments.   Indeed, the only way to relate physical magnitudes associated with different experiments is through the introduction of hidden variables, i.e., variables describing microscopic properties of the system that are not contained in the formalism of standard quantum theory. But superluminality would hardly be surprising in this case. Superluminal velocities are to be expected in any hidden-variables theory that reproduces the predictions of quantum mechanics, by virtue of Bell's theorem \cite{Bell}.

 For this reason, we believe that time-of-arrival experiments in tunneling systems could be of great significance for the foundations of quantum mechanics. We expect that the probability distribution Eq. (\ref{plt}) for the time of arrival is incompatible with any local hidden-variable theory, i.e., a hidden-variable theory that admits no superluminal propagation. Thus, experiments aiming to the confirmation of Eq. (\ref{plt}) could provide a new ground for testing the predictions of quantum theory Vs. local realism. They could demonstrate that quantum `non-locality' refers not only to the correlations between quantum subsystems, but also to the values of temporal observables in non-composite quantum systems.

\section{Conclusions}
In this work, we developed an approach to tunneling time based on the definition of probabilities for quantum temporal observables. The tunneling time is defined in terms of a probability distribution $P(L,t)$ associated with time-of-arrival experiments in a tunneling set-up. The probability distribution $P(L, t)$ contains all information about the temporal properties of the tunneling process.

We presented a methodology for constructing such probabilities, that applies to any quantum system. In particular, we applied this method to spinless relativistic particles described by a complex scalar quantum field. We derived a simple expression for the probability distribution $P(L,t)$, in which all information about the potential barrier is encoded in a complex coefficient $A_k$ constructed from the transmission and reflection coefficients of the potential, $T_k$ and $R_k$ respectively. We found that the total transmission probability is proportional to $|A_k|^2$ rather than to $|T_k|^2$ and that this difference may be very pronounced for highly asymmetric potentials.

We evaluated the probability distribution $P(L,t)$ explicitly for piecewise constant potentials. The case of the double barrier potential is of particular significance, because it provides an explicit demonstration that a single parameter (like the tunneling time) does not suffice to describe all temporal aspects of the tunneling process.

In the present paper, we specialized to the standard case of  a potential barrier that is defined in terms of a classical, background electric field. However, the applicability of the method is not restricted to this case. The probability distribution $P(L,t)$ is defined in terms of the two-point correlation function of any field theory with no restriction on the dynamics. Thus, the method presented here can be applied to a broader set of processes, for example, tunneling in open quantum systems (including the effect of dissipation and noise), or to incorporate the backreaction of the microscopic particles on the potential barrier.

\end{document}